\RequirePackage[fleqn]{amsmath}
\documentclass[12pt]{iopart}

\usepackage{graphicx}
\usepackage{amssymb,amsfonts,amsmath}
\usepackage{epstopdf}
\usepackage{color}
\usepackage{bm}
\usepackage[breaklinks=true,colorlinks=true,linkcolor=blue,urlcolor=blue,citecolor=blue]{hyperref}
\usepackage{cite}

\def\pa{\partial\Omega}
\def\T{{\mathcal T}}

\def\L{{\mathcal L}}
\def\R{{\mathbb R}}
\def\E{{\mathbb E}}
\def\M{{\mathcal M}}

\def\P{{\mathbb P}}

\def\s{\bm{s}}
\def\x{\bm{x}}

\def\diam{\textrm{diam}}

\def\erfc{\textrm{erfc}}

\def\pa{{\partial \Omega}}

\def\M{\mathcal{M}}
\def\P{\mathbb{P}}

\def\R{\mathbb{R}}

\def\T{\Gamma}
\def\TT{\mathcal{T}}

\def\I{\mathbb{I}}
\def\E{\mathbb{E}}

\def\X{\mathbf{X}}

\def\N{\mathcal{N}}
\def\S{\mathcal{S}}

\def\L{\mathcal{L}}

\def\ttau{\delta}

\begin{document}

\title[Statistics of diffusive encounters with a small target...]{Statistics of diffusive encounters with a small target: \\ Three complementary approaches}

\author{Denis~S.~Grebenkov}
\address{Laboratoire de Physique de la Mati\`{e}re Condens\'{e}e (UMR 7643), \\ 
CNRS -- Ecole Polytechnique, IP Paris, 91128 Palaiseau, France;}
\address{Institute for Physics and Astronomy, University of Potsdam, 14476 Potsdam-Golm,
Germany}
\ead{denis.grebenkov@polytechnique.edu}

\begin{abstract}
Diffusive search for a static target is a common problem in
statistical physics with numerous applications in chemistry and
biology.  We look at this problem from a different perspective and
investigate the statistics of encounters between the diffusing
particle and the target.  While an exact solution of this problem was
recently derived in the form of a spectral expansion over the
eigenbasis of the Dirichlet-to-Neumann operator, the latter is
generally difficult to access for an arbitrary target.  In this paper,
we present three complementary approaches to approximate the
probability density of the rescaled number of encounters with a small
target in a bounded confining domain.  In particular, we derive a
simple fully explicit approximation, which depends only on a few
geometric characteristics such as the surface area and the harmonic
capacity of the target, and the volume of the confining domain.  We
discuss the advantages and limitations of three approaches and check
their accuracy.  We also deduce an explicit approximation for the
distribution of the first-crossing time, at which the number of
encounters exceeds a prescribed threshold.  Its relations to common
first-passage time problems are discussed.
\end{abstract}






\date{\today}

\maketitle

\section{Introduction}

\begin{figure}
\begin{center}
\includegraphics[width=60mm]{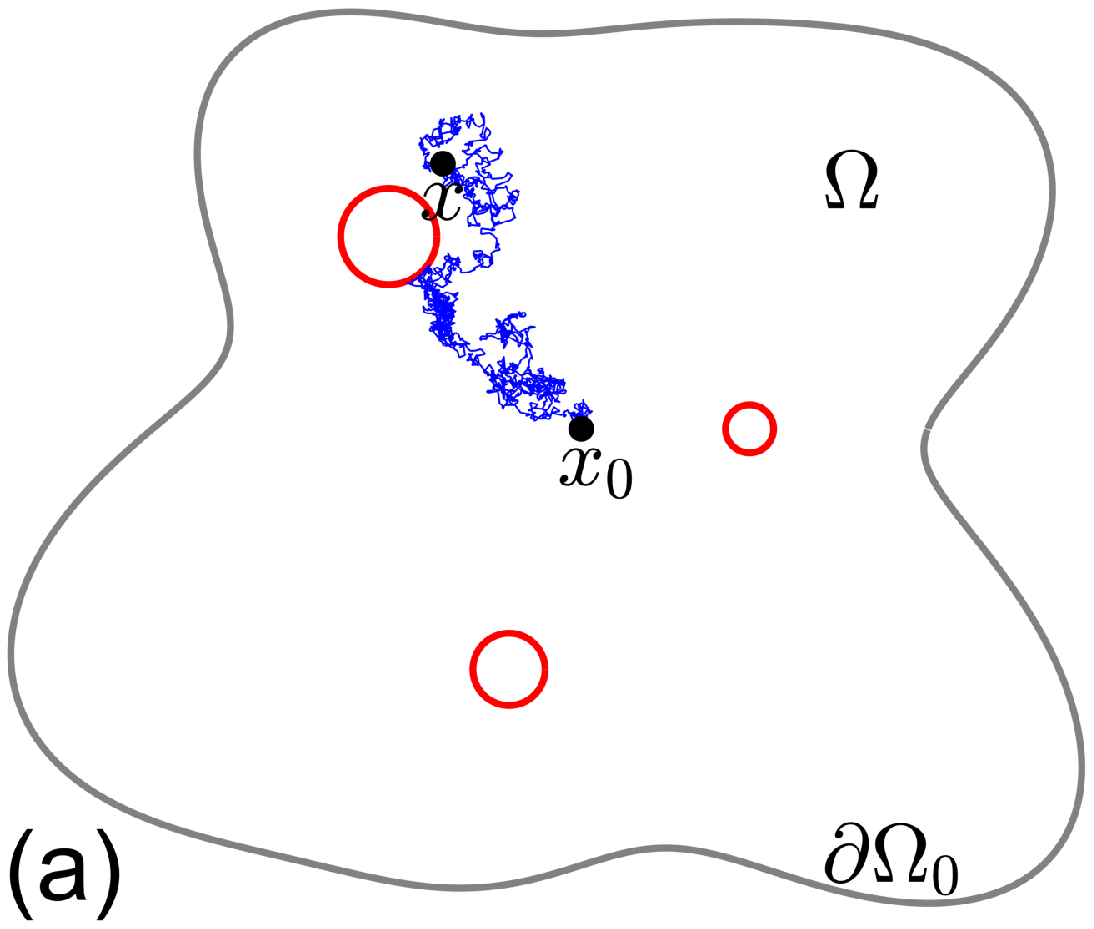} 
\includegraphics[width=60mm]{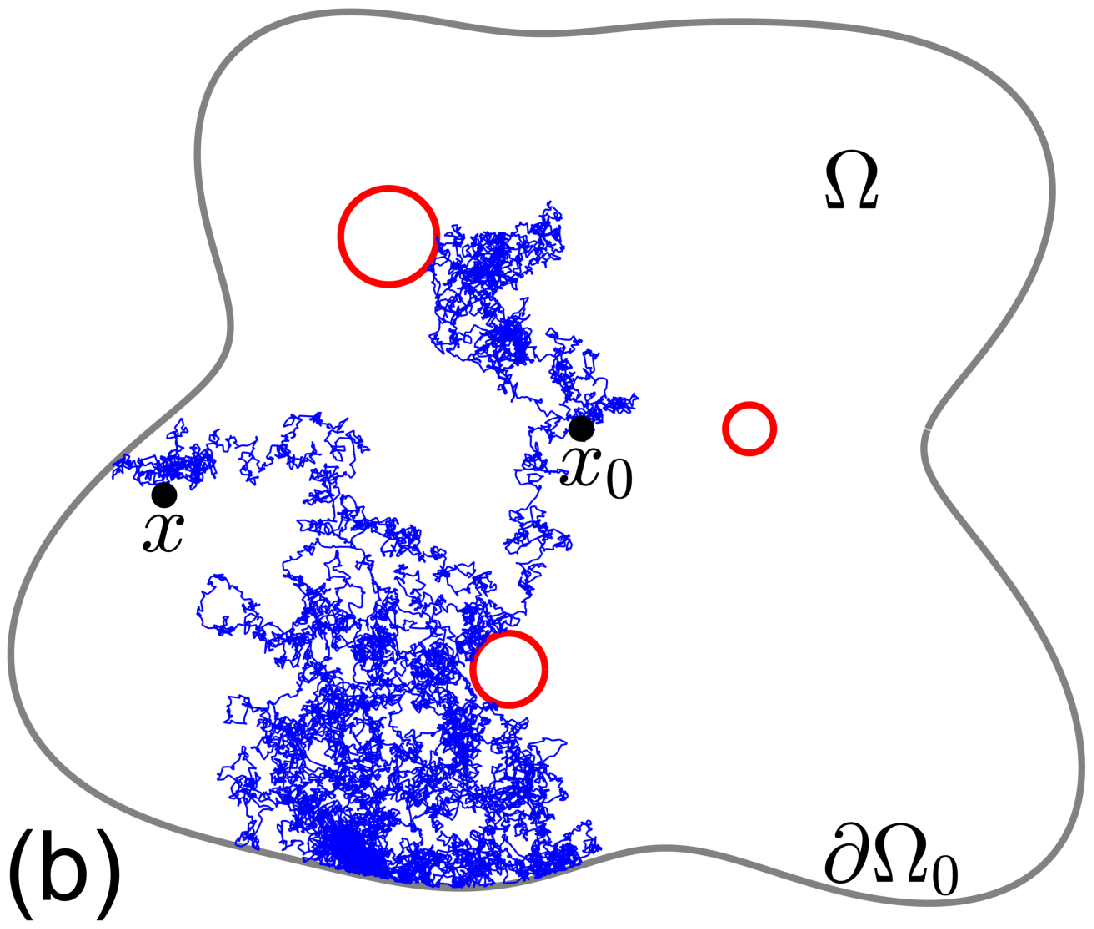} 
\end{center}
\caption{
Schematic illustration of a random trajectory (in blue) of a particle
that started from $\x_0$ and diffused inside a confining domain
$\Omega$ delimited by a reflecting boundary $\pa_0$ (in gray) towards
small targets (in red).  {\bf (a)} At short times, the particle
typically encounters either none, or one target.  {\bf (b)} At long
times, the particle can encounter many targets or realize several
``long-range returns'' to the same target. }
\label{fig:scheme}
\end{figure}

Various aspects of the first-passage problem for perfect and partially
reactive targets have been intensively studied over the past two
decades
\cite{Rice,Lauffenburger,Redner,Schuss,Metzler,Oshanin,Berg77,Weiss86,Condamin07,Grebenkov07,Benichou08,Benichou10,Benichou10b,Benichou11,Bressloff13,Benichou14,Godec16,Godec16b,Grebenkov16,Chechkin17,Lanoiselee18,Levernier19}.
A somewhat related but poorer understood problem is the statistics of
encounters of a diffusing particle with a target.  How many times does
the particle meet the target?  Conventionally, the statistics of
encounters and the related first-encounter times were studied for two
(or many) mobile particles that can represent, e.g., a protein and its
receptor, or a prey and its predator
\cite{Szabo88,Redner99,Bartumeus02,James08,Oshanin09,Sanders09,Tejedor11,Amitai12,Tzou14,Agliari14,Agliari16,Peng19,LeVot20,Nayak20,LeVot22}.
In turn, we are interested here in the number of encounters of a
diffusing particle with a static target up to time $t$.  For a random
walk on a lattice or a graph, this is the random number of visits of a
given target site (or a group of such sites) up to time $t$.  For
continuous diffusion, this number can be related to the residence time
of Brownian motion in a given subset of a confining domain.
Alternatively, one can consider the target located on an impenetrable
boundary (Fig. \ref{fig:scheme}), and the number of encounters is
directly related to the so-called boundary local time $\ell_t$ on that
region up to time $t$ \cite{Levy,Ito,Freidlin,Grebenkov07a}.  As
explained below, the random variable $\ell_t$ can be defined as a
rescaled residence time in a thin boundary layer near the target, or
as a rescaled number of downcrossings (encounters) of that layer.
The statistics of such encounters was recently shown to be tightly
related to the survival probability on a partially reactive target
\cite{Grebenkov19b,Grebenkov20a}; it can therefore be viewed as
a complementary insight onto the latter problem.  Moreover, the
knowledge of this statistics allows one to investigate much more
general surface-reaction mechanisms, for instance, those with
encounter-dependent reactivity \cite{Grebenkov20a}, which go far
beyond the common setting of partially reactive targets
\cite{Collins49,Sano79,Sano81,Shoup82,Sapoval94,Filoche99,Benichou00,Sapoval02,Grebenkov03,Grebenkov05,Grebenkov06a,Grebenkov06,Traytak07,Bressloff08,Singer08,Grebenkov10a,Grebenkov10b,Lawley15,Grebenkov15}.  A
general spectral representation of the distribution of encounters and
several explicit examples have been studied within the so-called
encounter-based approach
\cite{Grebenkov20c,Grebenkov20b,Grebenkov21a,Grebenkov22,Bressloff22b}.
However, to our knowledge, this problem has not been addressed in the
common case of a {\it small} target surrounded by an outer reflecting
boundary of a confining domain.

In this paper, we employ three recently developed approximations to
study the statistics of diffusive encounters with a small target (or
multiple targets).  In Sec. \ref{sec:localtime}, we formulate the
problem and recall the definition of the boundary local time as a
proxy of the number of encounters.  We also summarize the theoretical
ground needed to describe the statistics of encounters and its
relation to diffusion-controlled reactions.  Section \ref{sec:general}
presents three complementary approaches to deal with small targets.
The first approach is the matched asymptotic analysis (MAA), which was
systematically employed over the past three decades to investigate
various first-passage times
\cite{Mazya85,Ward93,Kolokolnikov05,Singer06a,Singer06b,Singer06c,Pillay10,Cheviakov10,Cheviakov11,Cheviakov12}
(see also a review \cite{Holcman14}).  Recently, Bressloff extended
this method to the encounter-based description of diffusion-mediated
surface phenomena and derived the asymptotic expansion of the
so-called full propagator \cite{Bressloff22}.  In
Sec. \ref{sec:matched}, we apply his asymptotic results to get the
statistics of diffusive encounters and discuss advantages and
limitations of this powerful method.  The second approach relies on an
explicit approximation for the principal eigenvalue of the Laplace
operator in the presence of a small partially reactive target
\cite{Chaigneau22}.  The smallness of the target also ensures that the
principal eigenmode provides the dominant contribution to the
volume-averaged survival probability, from which the distribution of
encounters will be deduced in a fully explicit way
(Sec. \ref{sec:eigenvalue}).  The derived approximation in
Eq. (\ref{eq:rho_ell}) is one of the main results of the paper.
The third approach is based on the self-consistent approximation
(SCA), which was originally proposed for computing reaction rates
\cite{Shoup81} and then extended for the analysis of first-passage
times
\cite{Grebenkov17a,Grebenkov17b,Grebenkov18a,Grebenkov19,Grebenkov21d}.
We employ its most general form given in \cite{Grebenkov21d} to deduce
an approximate spectral representation for the distribution of
diffusive encounters (Sec. \ref{sec:SCA}).  This formal representation
clarifies some general properties of the distribution.  Moreover, in
some symmetric domains such as, e.g., a spherical target surrounded by
a larger concentric reflecting sphere, the self-consistent
approximation turns out to be exact and can thus serve as a benchmark
for accessing the quality of two other approximations.  We choose this
geometric setting to illustrate the accuracy of different approaches
in Sec. \ref{sec:sphere}.  In Sec. \ref{sec:discussion}, we obtain an
explicit approximation (\ref{eq:Ut_PEA}) for the probability density
of the first-crossing time of a given threshold $\ell$ by $\ell_t$.
Section \ref{sec:conclusion} summarizes the main findings of the
paper, their applications and future perspectives.

\section{Boundary local time}
\label{sec:localtime}

We consider a point-like particle that undergoes reflected Brownian
motion with a constant diffusivity $D$ inside an Euclidean domain
$\Omega \subset \R^d$ with a smooth bounded impermeable boundary
$\pa$.  The particle starts from a point $\x_0 \in \Omega$ at time
$0$, and its random position at time $t$ is denoted as $\X_t$.  We are
interested in the statistics of encounters of the particle with a
subset $\T$ of the boundary that we call a target.  If the subset $\T$
is not connected, one can speak about multiple (disconnected) targets.
For the sake of clarity, we mainly speak about a connected (single)
target, even though our results are applicable to multiple targets if
they are located sufficiently far away from each other (see below).
The remaining part of the boundary, $\pa_0 = \pa \backslash \T$,
ensures normal reflections that confine the particle inside $\Omega$.
Following L\'evy's construction \cite{Levy}, we introduce the boundary
local time $\ell_t$ spent {\it on the target} up to time $t$.  This is
a stochastic process that can be understood as the renormalized
residence time of $\X_t$ in a thin layer of width $a$ near the target,
$\T_a = \{ \x \in \Omega~:~ |\x - \T| < a\}$, up time $t$
\cite{Ito,Freidlin}:
\begin{equation}  \label{eq:ell_res}
\ell_t = \lim\limits_{a \to 0} \frac{D}{a} \underbrace{\int\limits_0^t dt' \, \I_{\T_a}(\X_{t'})}_{\textrm{residence time in}~\T_a} ,
\end{equation}
where $\I_{\T_a}(\x)$ is the indicator function of $\T_a$:
$\I_{\T_a}(\x) = 1$ if $\x\in\T_a$, and $0$ otherwise.  This relation
highlights that the residence time in the layer $\T_a$ vanishes in the
limit $a\to 0$ when $\T_a$ shrinks to the target $\T$.  This is not
surprising given that the boundary $\pa$ has a lower dimension, $d-1$,
as compared to the dimension $d$ of the domain $\Omega$, and the
residence time on the boundary or any of its subsets is strictly zero.
In turn, the rescaling of the residence time in the layer $\T_a$ by
its width $a$ yields a well-defined limit -- the boundary local time.
Importantly, Eq. (\ref{eq:ell_res}) implies that the residence time in
a thin layer $\T_a$ can be approximated as $a \ell_t/D$, as soon as
$a$ is small enough.  The boundary local time $\ell_t$ is thus the
proper intrinsic characteristics of reflected Brownian motion on the
target, which is independent of the layer width used.  Note that
$\ell_t$ has units of length, while $\ell_t/D$ has units of time per
length.

The boundary local time $\ell_t$ is also related to the number
$\N_t^{a}$ of downcrossings of the boundary layer $\T_a$ by reflected
Brownian motion up to time $t$, multiplied by $a$, in the limit $a\to
0$ \cite{Ito,Freidlin}:
\begin{equation} \label{eq:ellt_N} 
\ell_t = \lim\limits_{a \to 0} a \, \N_t^{a} .
\end{equation}
The number of downcrossings can thus be interpreted as the number of
encounters of the particle with the target.  Expectedly, this number
diverges in the limit $a\to 0$, because Brownian motion is known to
hit a smooth boundary an infinite number of times during any time
period after the first hit \cite{Morters}.  A finite thickness $a$ of
the layer is needed here to overcome this problem and to get a proper
definition for the number of encounters (see further discussions in
\cite{Grebenkov19b,Grebenkov20a}).  As previously, $\ell_t/a$ is a
good proxy for the number of encounters when $a$ is small enough.  In
the following, we focus on the boundary local time $\ell_t$, sometimes
referring to it as the rescaled number of encounters.

The boundary local time plays the central role in the encounter-based
approach to surface-mediated diffusion phenomena
\cite{Grebenkov19b,Grebenkov20a,Grebenkov20b,Grebenkov20c,Grebenkov21a,Grebenkov22,Bressloff22b}.
In particular, the joint distribution of the particle position $\X_t$
and its boundary local time $\ell_t$ is characterized by the joint
probability density $P(\x,\ell,t|\x_0)$ that was called the full
propagator.  The integral of $P(\x,\ell,t|\x_0)$ over the arrival
point $\x$ yields the (marginal) probability density of the boundary
local time $\ell_t$:
\begin{equation}  \label{eq:rho_def}
\rho(\ell,t|\x_0) = \int\limits_{\Omega} d\x \, P(\x,\ell,t|\x_0).
\end{equation}
In turn, the Laplace transform of $P(\x,\ell,t|\x_0)$ with respect to
$\ell$ was shown to determine the conventional propagator
$G_q(\x,t|\x_0)$ in the presence of a partially reactive target
\cite{Grebenkov20a}:
\begin{equation}  \label{eq:Gq_rho0} 
G_q(\x,t|\x_0) = \int\limits_0^\infty d\ell \, e^{-q\ell} \, P(\x,\ell,t|\x_0) .
\end{equation}
The latter satisfies the diffusion equation
\begin{equation}  \label{eq:Gq_diff}
\partial_t G_q(\x,t|\x_0) = D \Delta G_q(\x,t|\x_0)  \quad (\x\in\Omega),
\end{equation}
subject to the initial condition $G_q(\x,0|\x_0) = \delta(\x-\x_0)$
with a Dirac distribution $\delta(\x-\x_0)$, and mixed Robin-Neumann
boundary condition:
\begin{subequations}  \label{eq:mixed}
\begin{align}   \label{eq:mixed1}
\partial_n G_q(\x,t|\x_0) + q G_q(\x,t|\x_0) & = 0 \quad (\x\in \T) , \\
\partial_n G_q(\x,t|\x_0) & = 0 \quad (\x\in \pa_0),
\end{align}
\end{subequations}
where $\Delta$ is the Laplace operator, and $\partial_n$ is the normal
derivative oriented outwards the domain $\Omega$.  This condition
claims that the diffusive flux $-D\partial_n G_q(\x,t|\x_0)$ vanishes
on the impermeable part of the boundary, $\pa_0$; in turn, on the
target $\T$, the diffusive flux is equal to the reaction flux, given
by $\kappa G_q(\x,t|\x_0)$, with a reactivity $\kappa = qD$.  In this
way, the parameter $q$ of the Laplace transform in
Eq. (\ref{eq:Gq_rho0}) naturally re-appears in the Robin boundary
condition (\ref{eq:mixed1}).  Expectedly, $q$ can vary between $0$ for
an inert target with Neumann boundary condition ($\partial_n
G_0(\x,t|\x_0)|_{\T} = 0$), and $+\infty$ for a perfectly reactive
target with Dirichlet boundary condition ($G_\infty(\x,t|\x_0)|_{\T} =
0$).
When the confining domain $\Omega$ is bounded, the propagator admits a
general spectral decomposition
\begin{equation}  \label{eq:Gq_spectral}
G_q(\x,t|\x_0) = \sum\limits_{k=1}^\infty  e^{-Dt\lambda_k^{(q)}} \, [u_k^{(q)}(\x)]^* \, u_k^{(q)}(\x_0)  ,
\end{equation}
where asterisk denotes complex conjugate, while $\lambda_k^{(q)}$ and
$u_k^{(q)}(\x)$ are the eigenvalues and eigenfunctions of the
(negative) Laplace operator in $\Omega$ with mixed Robin-Neumann
boundary condition:
\begin{subequations}
\begin{align}
\Delta u_k^{(q)} + \lambda_k^{(q)} u_k^{(q)} & = 0  \quad (\x\in\Omega) , \\  \label{eq:uk_Robin}
(\partial_n +q)u_k^{(q)} & = 0 \quad (\x\in\T), \\
\partial_n u_k^{(q)} & = 0 \quad (\x\in\pa_0) .
\end{align}
\end{subequations}
The eigenvalues are positive and indexed in an ascending order,
\begin{equation*}
0 < \lambda_1^{(q)} \leq \lambda_2^{(q)} \leq \ldots \leq \lambda_k^{(q)} \ldots \nearrow + \infty, 
\end{equation*}
while the eigenfunctions form a complete orthonormal basis in the
space $L_2(\Omega)$ of square-integrable functions on $\Omega$ (see a
review \cite{Grebenkov13} for other properties of Laplacian
eigenvalues and eigenfunctions).  The superscript $q$ highlights that
both $\lambda_k^{(q)}$ and $u_k^{(q)}(\x)$ depend {\it implicitly} on
the parameter $q$ via the Robin boundary condition
(\ref{eq:uk_Robin}).

As the integral of the propagator $G_q(\x,t|\x_0)$ over the arrival
point $\x$ yields the survival probability of the particle in the
presence of a partially reactive target,
\begin{equation}
S_q(t|\x_0) = \int\limits_{\Omega} d\x \, G_q(\x,t|\x_0),
\end{equation}
the latter is directly related to the probability density
$\rho(\ell,t|\x_0)$ of the boundary local time $\ell_t$:
\begin{equation}  \label{eq:Sq_rho0}  
S_q(t|\x_0) = \int\limits_0^\infty d\ell \, e^{-q\ell} \, \rho(\ell,t|\x_0) .
\end{equation}
We recall that $S_q(t|\x_0)$ satisfies the (backward) diffusion
equation
\begin{equation}
\partial_t S_q(t|\x_0) = D \Delta S_q(t|\x_0)  \quad (\x_0\in\Omega),
\end{equation}
subject to the initial (terminal) condition $S_q(0|\x_0) = 1$ and
mixed Robin-Neumann boundary condition:
\begin{subequations}  \label{eq:Sq_mixed}
\begin{align}
\partial_n S_q(t|\x_0) + q S_q(t|\x_0) & = 0 \quad (\x_0\in \T) , \\
\partial_n S_q(t|\x_0) & = 0 \quad (\x_0\in \pa_0).
\end{align}
\end{subequations}
According to Eq. (\ref{eq:Gq_spectral}), the survival probability
admits a general spectral decomposition
\begin{equation}
S_q(t|\x_0) = \sum\limits_{k=1}^\infty  e^{-Dt\lambda_k^{(q)}} \, [c_k^{(q)}]^* \, u_k^{(q)}(\x_0)  ,
\end{equation}
where
\begin{equation}    \label{eq:ck}
c_k^{(q)} = \int\limits_{\Omega} d\x \, u_k^{(q)}(\x) .
\end{equation}
The inverse Laplace transform of Eq. (\ref{eq:Sq_rho0}) with respect
to $q$ yields
\begin{equation}  \label{eq:rho_ell_Lq}
\rho(\ell,t|\x_0) = \L^{-1}_q \bigl\{ S_q(t|\x_0)\bigr\} 
= \L^{-1}_q \biggl\{\sum\limits_{k=1}^\infty e^{-Dt\lambda_k^{(q)}} \, [c_k^{(q)}]^* \, u_k^{(q)}(\x_0)  \biggr\} . 
\end{equation}

In the following, we often consider the volume-averaged case when the
starting point $\x_0$ is not fixed but uniformly distributed inside
the confining domain.  The corresponding survival probability $S_q(t)$
reads
\begin{equation}  \label{eq:Sq_volume}
S_q(t) = \frac{1}{|\Omega|} \int\limits_{\Omega} d\x_0 \, S_q(t|\x_0) = \sum\limits_{k=1}^\infty \frac{|c_k^{(q)}|^2}{|\Omega|} 
\, e^{-Dt\lambda_k^{(q)}}  \,,
\end{equation}
where $|\Omega|$ denotes the volume of the confining domain $\Omega$.
The relation (\ref{eq:Sq_rho0}) becomes
\begin{equation}  \label{eq:Sq_rho}
S_q(t) = \int\limits_0^\infty dq \, e^{-q\ell} \, \rho(\ell,t),
\end{equation}
where 
\begin{equation}
\rho(\ell,t) = \frac{1}{|\Omega|} \int\limits_{\Omega} d\x_0 \, \rho(\ell,t|\x_0) 
\end{equation}
is the probability density function of the boundary local time
$\ell_t$ for a particle started uniformly in the domain $\Omega$.

In the above setting, the bulk was not reactive, and the particle
could only react on the target $\T$.  If restricted diffusion occurs
inside a reactive medium, the particle has a finite lifetime due to
eventual disintegration, photobleaching or death during its motion
\cite{Yuste13,Meerson15,Grebenkov17c}.  In a common situation, the
lifetime $\tau$ of the particle is a random variable obeying the
exponential law, $\P\{\tau > t\} = e^{-pt}$, with $p$ being the decay
rate or, equivalently, $1/p$ being the mean lifetime.  In this case,
one can investigate the number of encounters with the target during
the particle's lifetime: $\ell_{\tau}$.  The probability density
function of this random variable can be easily obtained by averaging
$\rho(\ell,t|\x_0)$ over all possible times of particle's death:
\begin{equation}
\int\limits_0^\infty dt \, \underbrace{p\, e^{-pt}}_{\textrm{pdf of}~\tau} \, \rho(\ell,t|\x_0) = p \,\tilde{\rho}(\ell,p|\x_0),
\end{equation}
where $\tilde{\rho}(\ell,p|\x_0)$ is the Laplace transform of
$\rho(\ell,t|\x_0)$ with respect to $t$ (here and below, tilde denotes
Laplace-transformed quantities with respect to time $t$).  As a
consequence, Eq. (\ref{eq:Sq_rho0}) implies
\begin{equation}
\tilde{S}_q(p|\x_0) = \int\limits_0^\infty d\ell \, e^{-q\ell} \, \tilde{\rho}(\ell,p|\x_0) ,
\end{equation}
where $\tilde{S}_q(p|\x_0)$ satisfies the modified Helmholtz equation
\begin{equation}  \label{eq:Hp_Helmholtz}
(p - D\Delta) \tilde{S}(p|\x_0) = 1  \quad (\x_0\in\Omega),  
\end{equation}
subject to the mixed Robin-Neumann boundary condition:
\begin{subequations}  \label{eq:boundary3}
\begin{align}
\partial_n \tilde{S}_q(p|\x_0) + q \tilde{S}_q(p|\x_0) &= 0 \quad (\x_0\in\T), \\
\partial_n \tilde{S}_q(p|\x_0) & = 0 \quad (\x_0\in \pa_0). 
\end{align}
\end{subequations}

Finally, the Laplace-transformed probability density
$\tilde{\rho}(\ell,p|\x_0)$ admits another spectral expansion based on
the Dirichlet-to-Neumann operator $\M_p$ that associates to a given
function $f(\s)$ on the target $\T$ another function $g(\s)$ on that
target (see
\cite{Arendt14,Daners14,Elst14,Berhndt15,Arendt15,Hassell17,Girouard17}
for details):
\begin{equation}  \label{eq:DtN}
\M_p ~:~ f(\s) \to g(\s) = \left. \bigl(\partial_n \tilde{u}(\x)\bigr)\right|_{\x=\s\in \T}, 
\end{equation}
where $\tilde{u}(\x)$ satisfies
\begin{subequations}
\begin{align}  \label{eq:u_modified}
(p - D\Delta) \tilde{u}(\x) & =0 \quad (\x\in\Omega), \\ 
\tilde{u}(\x) & = f(\x) \quad (\x\in\T), \\ 
\partial_n \tilde{u}(\x) & = 0 \quad (\x\in\pa_0).  
\end{align}
\end{subequations}
It is known that $\M_p$ is a pseudo-differential self-adjoint
operator, whose positive eigenvalues $\mu_n^{(p)}$ can be ordered as
\begin{equation*}
0 \leq \mu_0^{(p)} \leq \mu_1^{(p)} \leq \ldots \leq \mu_n^{(p)} \leq \ldots \nearrow + \infty, 
\end{equation*}
while the associated eigenfunctions $v_n^{(p)}(\s)$ form a complete
orthonormal basis of $L_2(\T)$.  The superscript $p$ highlights that
both $\mu_n^{(p)}$ and $v_n^{(p)}(\s)$ depend {\it implicitly} on the
rate $p$ in Eq. (\ref{eq:u_modified}).  The following expansion was
derived in \cite{Grebenkov20a}
\begin{equation}
\tilde{P}(\x,\ell,p|\x_0) = \tilde{G}_\infty(\x,p|\x_0) \delta(\ell)   
 + \frac{1}{D} \sum\limits_{n=0}^\infty e^{-\mu_n^{(p)}\ell}\, [V_n^{(p)}(\x)]^* \, V_n^{(p)}(\x_0) ,
\end{equation}
where
\begin{equation} \label{eq:Vnp}
V_n^{(p)}(\x_0) = \int\limits_{\pa} d\s \, v_n^{(p)}(\s) \, \tilde{j}_\infty(\s,p|\x_0) 
\end{equation}
is the projection of the Laplace-transformed probability flux density
on the perfectly reactive target, $\tilde{j}_\infty(\s,p|\x_0) =
(-D\partial_n \tilde{G}_\infty(\x,p|\x_0))_{\x=\s}$, onto the
eigenfunction $v_n^{(p)}(\s)$.  According to Eq. (\ref{eq:rho_def}),
one has
\begin{equation}  \label{eq:rhop_exact}
\tilde{\rho}(\ell,p|\x_0) = \tilde{S}_\infty(p|\x_0) \delta(\ell)  
+ \sum\limits_{n=0}^\infty [C_n^{(p)}]^* \, \frac{\mu_n^{(p)}}{p} \, V_n^{(p)}(\x_0)\, e^{-\mu_n^{(p)}\ell} ,
\end{equation}
with
\begin{equation}  \label{eq:Cnp}
C_n^{(p)} = \int\limits_{\T} d\s \, v_n^{(p)}(\s),
\end{equation}
where we used the identity from \cite{Grebenkov19a}:
\begin{equation}  \label{eq:Vn_int}
\int\limits_{\Omega} d\x \, V_n^{(p)}(\x) = \frac{D}{p} \mu_n^{(p)} \int\limits_{\T} d\s \, v_n^{(p)}(\s) .
\end{equation}
The inverse Laplace transform of Eq. (\ref{eq:rhop_exact}) with
respect to $p$ formally reads
\begin{equation}   \label{eq:rho_exact}
\rho(\ell,t|\x_0) = S_\infty(t|\x_0) \delta(\ell) 
+ \L^{-1}_{p} \biggl\{\sum\limits_{n=0}^\infty [C_n^{(p)}]^* \, \frac{\mu_n^{(p)}}{p} \, V_n^{(p)}(\x_0) \, e^{-\mu_n^{(p)}\ell}  \biggr\}.
\end{equation}
The first term in Eq. (\ref{eq:rho_exact}) accounts for the
trajectories that have not encountered the target up to time $t$ (with
probability $S_\infty(t|\x_0)$) so that the associated boundary local
time remained zero.  In turn, the second term represents the
trajectories that have reached the target up to time $t$ and thus have
positive $\ell_t$.
Using again the identity (\ref{eq:Vn_int}), one can also treat the
case when the starting point $\x_0$ is uniformly distributed in
$\Omega$:
\begin{equation}   \label{eq:rho_exact_volume}
\rho(\ell,t) = S_\infty(t) \delta(\ell) 
 + \frac{D}{|\Omega|} \L^{-1}_{p} \biggl\{\sum\limits_{n=0}^\infty |C_n^{(p)}|^2 \, \frac{[\mu_n^{(p)}]^2}{p^2} \, e^{-\mu_n^{(p)}\ell}  \biggr\}.
\end{equation}

In summary, the probability density $\rho(\ell,t|\x_0)$ of the
boundary local time can be accessed in two complementary ways: either
via the inverse Laplace transform (\ref{eq:rho_ell_Lq}) with respect
to $q$, or via the inverse Laplace transform (\ref{eq:rho_exact}) with
respect to $p$.  These equivalent ways reflect the duality of bulk and
surface reaction mechanisms elaborated in \cite{Grebenkov20a}.  We
will employ both ways in the analysis of small targets.

\section{Three approaches}
\label{sec:general}

This section describes our main theoretical results on the statistics
of the boundary local time on a small target.  We present three
complementary approaches to address this problem.  We start in
Sec. \ref{sec:matched} by the matched asymptotic analysis, which aims
to match two approximate solutions -- an inner solution in the
vicinity of the target as if the target was located in the free space
(without confinement), and an outer solution as if the target was
point-like.  Our derivation relies on the asymptotic expansion of the
Laplace-transformed full propagator $\tilde{P}(\x,\ell,p|\x_0)$ that
was recently obtained by Bressloff \cite{Bressloff22}.  We show the
advantages and practical limitations of this general and powerful
technique.  In Sec. \ref{sec:eigenvalue}, we use a different strategy
based on an approximation for the principal eigenvalue of the Laplace
operator.  In this way, we derive a very simple yet accurate
approximation for the probability density of the boundary local time.
In Sec. \ref{sec:SCA}, we employ yet another approach relying on the
self-consistent approximation.  These three approaches provide
complemenary views onto the statistics of encounters with a small
target.

\subsection{Matched asymptotic analysis (MAA)} 
\label{sec:matched}

Bressloff developed the matched asymptotic analysis for the
Laplace-transformed full propagator $\tilde{P}(\x,\ell,p|\x_0)$ in
three dimensions \cite{Bressloff22}.  He considered a configuration of
$N$ spherical targets located at points $\x_1,\ldots,\x_N \in \Omega$
and having small radii $r_1,\ldots,r_N$ (an extension to nonspherical
targets was also discussed).  The targets were supposed to be located
far away from each other:
\begin{equation}  \label{eq:cond_separated}
\max_j\{ r_j \} \ll \min_{i\ne j} \{ |\x_i - \x_j|\} \,.
\end{equation}
In other words, if $R_{\rm min}$ is the minimal separation distance
between targets (the right-hand side), then $r_j/R_{\rm min} =
O(\epsilon)$, where $\epsilon$ is a small parameter.

In the leading-order term, Bressloff obtained a very simple relation,
\begin{equation}  \label{eq:tildeP_asympt}
\tilde{P}(\x,\ell,p|\x_0) = \tilde{G}_\infty(\x,p|\x_0) \delta(\ell) + \tilde{u}_0(\x,\ell,p|\x_0) + O(\epsilon),
\end{equation}
where $\tilde{G}_\infty(\x,p|\x_0)$ is the Laplace transform of
the propagator $G_\infty(\x,t|\x_0)$ defined by Eq. (\ref{eq:Gq_diff},
\ref{eq:mixed}) with $q = \infty$ (i.e., a perfectly reactive target with
Dirichlet boundary condition), and
\begin{equation}  \label{eq:tildeU0}
\tilde{u}_0(\x,\ell,p|\x_0) = 4\pi D \sum\limits_{j=1}^N e^{-\ell/r_j} \, \tilde{g}(\x_j,p|\x_0) \, \tilde{g}(\x,p|\x_j),
\end{equation}
with
\begin{equation}
\tilde{g}(\x,p|\x_0) = \frac{e^{-|\x-\x_0|\sqrt{p/D}}}{4\pi D |\x-\x_0|}
\end{equation}
being the fundamental solution of the modified Helmholtz equation in
$\R^3$:
\begin{equation}
(p - D \Delta) \tilde{g}(\x,p|\x_0) = \delta(\x-\x_0).
\end{equation}
The next-order term in Eq. (\ref{eq:tildeP_asympt}) was also
determined in \cite{Bressloff22} but its expression is more
sophisticated.

A probabilistic interpretation of Eq. (\ref{eq:tildeP_asympt}) is
instructive.  The first term represents the contribution of
trajectories that have not encountered any target so that the boundary
local time $\ell_t$ remained zero.  In turn, the second term describes
the trajectories that reached one of the targets.  As the targets are
small and well separated, their contributions are independent from
each other and thus additive.  For the $j$-th target, the factor
$\tilde{g}(\x_j,p|\x_0)$ describes the passage from the starting point
$\x_0$ to the target location $\x_j$, the factor $e^{-\ell/r_j}$
characterizes the acquired boundary local time, and the factor
$\tilde{g}(\x,p|\x_j)$ describes the motion from $\x_j$ to $\x$
(Fig. \ref{fig:scheme}(a)).  We recall that the inverse Laplace
transform of this factor,
\begin{equation}
g(\x_j,t|\x_0) = \frac{1}{(4\pi Dt)^{3/2}} \exp\biggl(-\frac{|\x_j-\x_0|^2}{4Dt}\biggr) \,,
\end{equation}
is the propagator of Brownian motion in the three-dimensional space
(without any target).  Naturally, the next-order terms account for
trajectories that have visited two or many targets
(Fig. \ref{fig:scheme}(b)).

This intuitive picture suggests that the leading-order asymptotic
expansion (\ref{eq:tildeP_asympt}) is expected to be most accurate for
large $p$; in fact, thinking of $p$ as the bulk reaction rate, one
deals here with a highly reactive medium, in which diffusion between
distant points is penalized: the contribution of rare trajectories
that visit two or many targets within the particle's lifetime (the
$O(\epsilon)$ term in Eq. (\ref{eq:tildeP_asympt})) is negligible.
Similarly, the possibility of a long excursion started from a single
target and returned to it, is also unlikely.  In other words, such
``long-range returns'' to the target are statistically suppressed.  We
can therefore anticipate that the following asymptotic results would
be accurate at short time.  In contrast, the contribution
$O(\epsilon)$ is expected to be relevant in the opposite limit $p\to
0$ or, equivalently, at long times.  We will come back to these
statements in Sec. \ref{sec:sphere}.

If the limitation to small $p$ can be ignored (i.e., if
Eq. (\ref{eq:tildeU0}) can be used for any $p$), the inverse Laplace
transform of Eq. (\ref{eq:tildeP_asympt}) implies
\begin{equation}  \label{eq:P_asympt}
P(\x,\ell,t|\x_0) = G_\infty(\x,t|\x_0) \delta(\ell) + u_0(\x,\ell,t|\x_0) + O(\epsilon),
\end{equation}
with
\begin{equation}
\fl
u_0(\x,\ell,t|\x_0) = \frac{1}{4\pi D} \sum\limits_{j=1}^N  \, \frac{e^{-\ell/r_j}}{|\x-\x_j| \, |\x_0-\x_j|}  
\, h(t,|\x-\x_j|+|\x_0-\x_j|) \,,
\end{equation}
where
\begin{equation}
h(t,x) = \frac{x \, e^{-x^2/(4Dt)}}{\sqrt{4\pi Dt^3}}
\end{equation}
is the L\'evy-Smirnov probability density of the first-passage time to
the origin for a one-dimensional Brownian motion started from $x$.
Substituting the asymptotic expansion (\ref{eq:P_asympt}) into
Eq. (\ref{eq:rho_def}), we get
\begin{equation}   \label{eq:rho_Bressloff}
\rho(\ell,t|\x_0) = S_\infty(t|\x_0) \delta(\ell) + u_0(\ell,t|\x_0) + O(\epsilon),
\end{equation}
where
\begin{equation}
u_0(\ell,t|\x_0) = \int\limits_\Omega d\x \, u_0(\x,\ell,t|\x_0) .
\end{equation}
In order to evaluate this contribution, one can look again at the
Laplace-transformed quantity $\tilde{u}_0(\x,\ell,p|\x_0)$.  If $p$ is
not too small (say, $\sqrt{p/D} \, \diam\{\Omega\} \gg 1$), the
integral of $\tilde{g}(\x,p|\x_j)$ over $\Omega$ can be accurately
approximated by integrating over the whole space $\R^3$:
\begin{equation}
\int\limits_{\Omega} d\x \, \tilde{g}(\x,p|\x_j) \approx \int\limits_{\R^3} d\x \, \tilde{g}(\x,p|\x_j) = \frac{1}{p} \,.
\end{equation}
As a consequence,
\begin{equation}
\tilde{u}_0(\ell,p|\x_0) \approx  4\pi D \sum\limits_{j=1}^N e^{-\ell/r_j} \, \tilde{g}(\x_j,p|\x_0) \, \frac{1}{p} \,,
\end{equation}
from which
\begin{equation}  \label{eq:u0x0}
u_0(\ell,t|\x_0) \approx \sum\limits_{j=1}^N \frac{e^{-\ell/r_j}}{|\x_j - \x_0|}\,  \erfc\biggl(\frac{|\x_j-\x_0|}{\sqrt{4Dt}}\biggr) ,
\end{equation}
where $\erfc(z)$ is the complementary error function.  This function
determines a fully explicit approximation (\ref{eq:rho_Bressloff}) to
the probability density $\rho(\ell,t|\x_0)$.

Moreover, if the starting point is uniformly distributed over the
confining domain $\Omega$, one has
\begin{equation}  \label{eq:rho_av_Bressloff}
\rho(\ell,t) = S_\infty(t) \delta(\ell) + u_0(\ell,t) + O(\epsilon),
\end{equation}
where
\begin{equation}
u_0(\ell,t) = \frac{1}{|\Omega|} \int\limits_{\Omega} d\x_0 \, u_0(\ell,t|\x_0) .
\end{equation}
As $u_0(\ell,t|\x_0)$ in Eq. (\ref{eq:u0x0}) is given as a sum, one
can compute $u_0(\ell,t)$ by integrating separately each term.  For
the $j$-th term, one can introduce local spherical coordinates
centered at $\x_j$ and ignore the presence of other targets due to the
well-separation condition (\ref{eq:cond_separated}).  In addition, the
earlier assumption of large $p$ is equivalent to considering the
short-time limit, in which the upper limit of the integral can be
replaced by infinity:
\begin{align}  \nonumber
u_0(\ell,t) & \approx \frac{1}{|\Omega|} \sum\limits_{j=1}^N \int\limits_{\Omega} d\x_0
\frac{e^{-\ell/r_j}}{|\x_j - \x_0|}\,  \erfc\biggl(\frac{|\x_j-\x_0|}{\sqrt{4Dt}}\biggr)  \\  \label{eq:U0_short}
& \approx \frac{1}{|\Omega|} \sum\limits_{j=1}^N e^{-\ell/r_j} \, 4\pi \int\limits_0^\infty dr \, r^2
\frac{1}{r}\,  \erfc\biggl(\frac{r}{\sqrt{4Dt}}\biggr)  
= \frac{4\pi Dt}{|\Omega|} \sum\limits_{j=1}^N e^{-\ell/r_j} .
\end{align}

For a single target or for multiple {\it identical} targets (with
equal radii $r_j = R$), the leading term of the asymptotic expansions
(\ref{eq:P_asympt}, \ref{eq:rho_Bressloff}, \ref{eq:rho_av_Bressloff})
exhibits the same dependence on $\ell$ via the factor $e^{-\ell/R}$.
As a consequence, the leading-order expansion can be interpreted as an
exponential distribution of the boundary local time $\ell_t$, to which
a probability measure atom at $\ell = 0$ is added:
\begin{equation}  \label{eq:rho_exp0}
\fl
\rho(\ell,t|\x_0) \approx S_\infty(t|\x_0) \delta(\ell) + e^{-\ell/R}  \, 
\sum\limits_{j=1}^N \frac{1}{|\x_j - \x_0|}\,  \erfc\biggl(\frac{|\x_j-\x_0|}{\sqrt{4Dt}}\biggr) + O(\epsilon),
\end{equation}
where we used Eq. (\ref{eq:u0x0}).  Similarly, Eq. (\ref{eq:U0_short})
yields for the uniformly distributed starting point:
\begin{equation}   \label{eq:rho_exp_MAA}
\rho(\ell,t) \approx S_\infty(t) \, \delta(\ell) + \frac{4\pi Dt}{|\Omega|} e^{-\ell/R} \,.
\end{equation}
As the probability densities $\rho(\ell,t|\x_0)$ and $\rho(\ell,t)$
must be normalized to $1$, the time-dependent prefactors in front of
$\delta(\ell)$ and $e^{-\ell/R}$ should be related as
\begin{subequations}  \label{eq:MMA_conditions}
\begin{align}
1 & = \int\limits_0^\infty d\ell \, \rho(\ell,t|\x_0) = S_\infty(t|\x_0)
 + R \sum\limits_{j=1}^N \frac{1}{|\x_j - \x_0|}\,  \erfc\biggl(\frac{|\x_j-\x_0|}{\sqrt{4Dt}}\biggr) , \\   \label{eq:cond_MMA}
1 & = \int\limits_0^\infty d\ell \, \rho(\ell,t) = S_\infty(t) + R \frac{4\pi Dt}{|\Omega|} \,.
\end{align}
\end{subequations}
The accuracy of these relations can serve as an indicator of the
quality of the exponential-like approximations (\ref{eq:rho_exp0},
\ref{eq:rho_exp_MAA}).  As we will discuss in Sec. \ref{sec:sphere},
these relations can be fulfilled for some intermediate range of times
but fail in both limits of short and long times.  This failure reveals
practical limitations of Eqs. (\ref{eq:rho_exp0},
\ref{eq:rho_exp_MAA}).  In turn, when the relations 
(\ref{eq:MMA_conditions}) are valid, one can replace
Eqs. (\ref{eq:rho_exp0}, \ref{eq:rho_exp_MAA}) by their equivalent
forms, which automatically respect the normalization:
\begin{equation}  \label{eq:rho_exp}
\rho(\ell,t|\x_0) \approx S_\infty(t|\x_0) \delta(\ell) + (1- S_\infty(t|\x_0)) \frac{e^{-\ell/R}}{R}  \, 
\end{equation}
and
\begin{equation}   \label{eq:rho_exp_av}
\rho(\ell,t) \approx S_\infty(t) \, \delta(\ell) + (1 - S_\infty(t)) \frac{e^{-\ell/R}}{R} \,,
\end{equation}
with $S_\infty(t)$ given by Eq. (\ref{eq:Sq_volume}). 
We will discuss the validity of these exponential-like distributions
in Sec. \ref{sec:sphere}.

In summary, one sees that the matched asymptotic analysis is a general
and powerful technique to access the asymptotic behavior of the full
propagator and the related quantities.  While the leading-order term
of the regular part of the Laplace-transformed full propagator,
$\tilde{u}_0(\x,\ell,p|\x_0)$, admits a simple probabilistic
interpretation, its form in time domain is less intuitive.  Moreover,
getting explicit approximations for the related quantities such as the
probability density of the boundary local time, requires further
simplifying assumptions that may limit the range of their
applicability.  Most importantly, the next-order terms accounting for
the contribution of trajectories visiting several targets (or
performing several ``long-range returns'' to a single target), may
become relevant at long times.  Even though the matched asymptotic
analysis offers a systematic way to access these terms, their
derivation and dependence on the parameters become much more
sophisticated.  For this reason, one may search for alternative
approaches to get an approximation of the probability density of the
boundary local time on a small target.

\subsection{Principal eigenvalue approximation (PEA)}
\label{sec:eigenvalue}

When the target is small, the ground eigenfunction of the Laplace
operator is nearly constant, $u_1^{(q)}(\x)\approx |\Omega|^{-1/2}$,
except for a vicinity of the target (see \cite{Chaigneau22} and
references therein).  As a consequence, the coefficient $c_k^{(q)}$ in
Eq. (\ref{eq:ck}), in which the eigenfuction $u_k^{(q)}$ is projected
onto a constant and thus onto $u_1^{(q)}$, can be approximated as
$c_k^{(q)} \approx \delta_{k,1}$, where $\delta_{k,1}$ is the
Kronecker symbol: $\delta_{k,1} = 1$ for $k = 1$ and $0$ otherwise.
The ground eigenmode provides therefore the major contribution to the
volume-averaged survival probability:
\begin{equation}  \label{eq:Sq_approx}
S_q(t) \approx e^{-Dt\lambda_1^{(q)}}  .
\end{equation}
If the small target $\T$ is located far away from the reflecting
boundary $\pa_0$ of a bounded domain in $\R^d$ with $d \geq 3$, the
principal eigenvalue can be approximated as \cite{Chaigneau22}:
\begin{equation}  \label{eq:lambda1}
\lambda_1^{(q)} \approx \lambda_1^{(\infty)} \, \frac{qL}{1 + qL} \,,  \qquad \lambda_1^{(\infty)} \approx \frac{C}{|\Omega|} \,.
\end{equation}
Here $L = |\T|/C$ was called the trapping length of the target, with
$|\T|$ and $C$ being respectively the surface area and the (harmonic)
capacity of the target.  The eigenvalue $\lambda_1^{(\infty)}$
corresponds to a perfectly reactive target ($q = \infty$) with
Dirichlet boundary condition.  The validity and high accuracy of the
principal eigenvalue approximation (\ref{eq:lambda1}) were confirmed
for anisotropic targets in $\R^d$ with several values of $d \geq 3$
\cite{Chaigneau22}.  Actually, the approximation (\ref{eq:lambda1})
was getting more and more accurate as $d$ increases.  In turn, one may
need to include the known first-order correction to the capacity $C$
in three dimensions (see \cite{Cheviakov11,Chaigneau22} and references
therein, as well as Eq. (\ref{eq:Cprime}) below).  Even though the
analysis in \cite{Chaigneau22} was focused on a single target, the
derivation remains applicable for multiple, well-separated small
targets satisfying the condition (\ref{eq:cond_separated}).  In this
case, $C$ is the sum of harmonic capacities of all targets, while
$|\Gamma|$ is the sum of their surface areas.

Substituting the approximation (\ref{eq:lambda1}) into
Eq. (\ref{eq:Sq_approx}), one can invert the Laplace transform in
Eq. (\ref{eq:Sq_rho}) to get
\begin{equation}
\rho(\ell,t) \approx  \rho_{\rm PEA}(\ell,t) ,
\end{equation}
where
\begin{align} \nonumber
\rho_{\rm PEA}(\ell,t) &= \L^{-1}_q\biggl\{ \exp\biggl(-Dt \lambda_1^{(\infty)} \frac{qL}{1 + qL} \biggr) \biggr \}  \\  \nonumber
& = e^{- Dt\lambda_1^{(\infty)}} \biggl(\delta(\ell) 
+ \sum\limits_{n=1}^\infty \frac{(Dt\lambda_1^{(\infty)})^n}{n!} \L^{-1}_q\{ (1+qL)^{-n} \} \biggr) \\  \label{eq:rho_series}
& = e^{- Dt\lambda_1^{(\infty)}} \biggl(\delta(\ell) 
+ \frac{Dt\lambda_1^{(\infty)}}{L} e^{-\ell/L} \sum\limits_{n=0}^\infty \frac{(Dt\lambda_1^{(\infty)}\ell/L)^n}{n! (n+1)!} \biggr),
\end{align}
which can also be written as
\begin{equation}  \label{eq:rho_ell}
\rho_{\rm PEA}(\ell,t) = e^{-t/T} \delta(\ell) + \sqrt{\frac{t/T}{\ell L}} 
e^{-\ell/L - t/T}  I_1\biggl(2\sqrt{(t/T)(\ell/L)}\biggr) ,
\end{equation}
with $I_\nu(z)$ being the modified Bessel function of the first kind,
and $T = 1/(D\lambda_1^{(\infty)})$.  This is one of the main results
of the paper.  As the factor in front of $\delta(\ell)$ is an
approximation of the survival probability $S_\infty(t)$, the first
term is again interpreted as the contribution of trajectories that
never reached the target up to time $t$ (and thus $\ell_t = 0$).  In
turn, the second term accounts for the trajectories that arrived onto
the target and thus yield positive $\ell_t$.  Despite its approximate
character, this probability density is correctly normalized for any
$t$:
\begin{equation*}
\int\limits_0^\infty d\ell \, \rho_{\rm PEA}(\ell,t) = 1 .
\end{equation*}
When $Dt\lambda_1^{(\infty)}\ell/L \gg 1$, the asymptotic behavior of
$I_1(z)$ yields
\begin{equation}    \label{eq:rho_ell_app}
\rho_{\rm PEA}(\ell,t) \approx e^{-t/T} \delta(\ell) 
+ \frac{(t/T)^{\frac14}}{2\sqrt{\pi} \, (\ell/L)^{\frac34} L}
\exp\biggl(-\biggl(\sqrt{\ell/L} - \sqrt{t/T}\biggr)^2\biggr) .
\end{equation}
The moments of $\ell_t$ can be easily deduced from
Eq. (\ref{eq:rho_ell}):
\begin{align}  \nonumber
\E\{ [\ell_t]^k \} & = \int\limits_0^\infty d\ell \, \ell^k \, \rho(\ell,t) \approx \sqrt{t/T} e^{-t/T} L^k
\int\limits_0^\infty dz \, z^{k-1/2} \, e^{-z} \, I_1\bigl(2\sqrt{(t/T)z}\bigr) \\
& = (t/T)\, L^k\,  k!\, M\bigl(1-k;2; -t/T\bigr) ,
\end{align}
where $M(a;b;z)$ is the Kummer's confluent hypergeometric function.
One sees that for any positive integer $k$, the right-hand side is a
polynomial of $t/T$ of order $k-1$.  For instance, one gets
\begin{equation}
\E\{ \ell_t \} \approx L (t/T) \approx Dt |\T|/|\Omega| , \qquad
\mathrm{var}\{ \ell_t\} \approx 2L^2 (t/T) , 
\end{equation}
in agreement with general results \cite{Grebenkov19b}.

The accuracy of this approximation will be discussed in
Sec. \ref{sec:sphere}.  We stress that the approximation
(\ref{eq:rho_ell}) is fully explicit and includes just few geometric
parameters: the surface area and the capacity of the target, as well
as the volume of the domain.  In the case of a spherical target of
radius $R$, one has $|\T| = 4\pi R^2$ and $C = 4\pi R$, i.e., the
trapping length is simply $L = R$.  However, there is no restriction
neither on the shape on the target, nor on the space dimensionality $d
\geq 3$ (see the discussion of the planar case in
Sec. \ref{sec:conclusion}).  In other words, the approximation is
valid for a general confining domain and any small enough target (up
to some mathematical restrictions, e.g., on boundary smoothness for a
rigorous formulation of the problem).

It is instructive to compare the regular part of the approximation
(\ref{eq:rho_ell}) to the exponential-like behavior
(\ref{eq:rho_exp_av}) for a single spherical particle with $L = R$.
Despite the common factor $e^{-\ell/R}$, the dependence on $\ell$ is
more sophisticated in Eq. (\ref{eq:rho_ell}).  In fact, $\ell$ appears
in the argument of the modified Bessel function and is thus coupled to
time $t$, which thus controls the behavior of $I_1(z)$.  This can also
be seen in the long-time asymptotic relation (\ref{eq:rho_ell_app}),
which exhibits a distinct maximum of the boundary local time $\ell_t$
around the mean value $(t/T)L$ that grows with time.  In contrast, the
mean value predicted by Eq. (\ref{eq:rho_exp_av}), $R(1 -
S_{\infty}(t))$, approaches a constant $R$ as time $t$ grows.  This
behavior, which contradicts the general properties of the boundary
local time in confined domains, is in turn reminiscent to the case of
a spherical target in $\R^3$, as if the reflecting boundary $\pa_0$
was moved to infinity \cite{Grebenkov19b}.  Once again, this
discrepancy highlights the limitations of the matched asymptotic
expansion at long times.

\subsection{Self-consistent approximation (SCA)}
\label{sec:SCA}

Even though the principal eigenvalue approximation provides a simple
form of the probability density $\rho(\ell,t)$, it is instructive to
discuss yet another approach to this problem.  In
\cite{Grebenkov17a,Grebenkov17b,Grebenkov18a,Grebenkov19,Grebenkov21d},
a self-consistent approximation was developed to calculate the
survival probability $S_q(t|\x_0)$, in which the mixed boundary
condition (\ref{eq:boundary3}) was replaced by an effective
inhomogeneous Neumann boundary condition, with a constant flux on the
target.  For a small target in a general confining domain, the
self-consistent approximation reads \cite{Grebenkov21d}
\begin{equation}  \label{eq:SCA}
\tilde{S}_q^{\rm app}(p|\x_0) = \frac{1}{p} - \frac{1}{p} \biggl(\frac{1}{q} + \frac{1}{Q_p} \biggr)^{-1} \, F_p(\x_0),
\end{equation}
where 
\begin{equation}  \label{eq:Qp}
\frac{1}{Q_p} = \frac{1}{|\T|} \sum\limits_{n=0}^\infty \frac{|C_n^{(p)}|^2}{\mu_n^{(p)}} 
\end{equation}
and
\begin{equation}  \label{eq:Fx}
F_p(\x_0) = \sum\limits_{n=0}^\infty \frac{[C_n^{(p)}]^*}{\mu_n^{(p)}} V_n^{(p)}(\x_0)  ,
\end{equation}
with $V_n^{(p)}(\x_0)$ and $C_n^{(p)}$ being given by
Eqs. (\ref{eq:Vnp}, \ref{eq:Cnp}).  The inversion of the Laplace
transform in Eq. (\ref{eq:Sq_rho0}) with respect to $q$ yields
\begin{equation}   \label{eq:Pell}
\tilde{\rho}^{\rm app}(\ell,p|\x_0) = \frac{1 - Q_p F_p(\x_0)}{p}\, \delta(\ell) + \frac{Q_p^2 \, F_p(\x_0)}{p} \, \exp(-Q_p \ell) .
\end{equation}
As previously, the first term represents the contribution of
trajectories that do not encounter the target, whereas the second term
gives the contribution of trajectories that encountered the target and
thus increased the boundary local time.  One can easily check that the
approximate probability density in Eq. (\ref{eq:Pell}) is correctly
normalized:
\begin{equation}
\int\limits_0^\infty d\ell  \, \tilde{\rho}^{\rm app}(\ell,p|\x_0) = \frac{1}{p} \,.
\end{equation}
Recalling that $p \, \tilde{\rho}(\ell,p|\x_0)$ is the probability
density of the boundary local time $\ell_\tau$ acquired until the
particle's death, one realizes that Eq. (\ref{eq:Pell}) provides again
a simple exponential-like distribution of $\ell_\tau$, whose
parameters are determined by $Q_p$ and $F_p(\x_0)$.  We stress that
the above description has no restriction on the connectivity of the
target $\Gamma$, i.e., it is applicable to multiple targets as well.

While the probability density (\ref{eq:Pell}) has a simple form in the
Laplace domain, its parameters $Q_p$ and $F_p(\x_0)$ depend on the
confining domain and the target in a sophisticated way (via the
spectral properties of the Dirichlet-to-Neumann operator).  In
addition, the inverse Laplace transform with respect to $p$ is still
needed to access the probability density function $\rho(\ell,t|\x_0)$
of the boundary local time $\ell_t$.  As a consequence, this approach
may look too sophisticated and less informative as compared to the
former ones discussed in Sec. \ref{sec:matched} and
\ref{sec:eigenvalue}.  At the same time, it brings complementary
insights onto the statistics of diffusive encounters, as described
below.  In addition, the self-consistent approximation becomes exact
in some symmetric domains and can thus be used as a benchmark for
validating the accuracy of other approximations (see
Sec. \ref{sec:sphere}).  Finally, the self-consistent approximation
does not rely on a sufficient separation between the target and the
outer boundary (that was required for the principal value
approximation in Sec. \ref{sec:eigenvalue}), nor on a sufficient
separation between targets (that was required for the matched
asymptotic analysis in Sec. \ref{sec:matched}).  In other words, the
underlying assumptions are weaker than in two other cases.

When the starting point $\x_0$ is uniformly distributed inside the
confining domain $\Omega$, the volume average reads:
\begin{equation}  \label{eq:rho_app0}
\tilde{\rho}^{\rm app}(\ell,p) = \frac{1}{|\Omega|} \int\limits_{\Omega} d\x_0 \, \tilde{\rho}^{\rm app}(\ell,p|\x_0) 
 = \frac{1 - Q_p \overline{F_p}}{p} \delta(\ell) + \frac{Q_p^2 \, \overline{F_p}}{p} \, e^{- Q_p \ell} ,
\end{equation}
with
\begin{equation*}
\fl
\overline{F_p} = \frac{1}{|\Omega|} \int\limits_{\Omega} d\x_0 \,F_p(\x_0) 
 = \frac{1}{|\Omega|} \sum\limits_{n=0}^{\infty} \frac{[C_n^{(p)}]^*}{\mu_n^{(p)}} \int\limits_{\Omega} d\x_0 \, V_n^{(p)}(\x_0) 
 =  \frac{D}{p|\Omega|} \sum\limits_{n=0}^{\infty} |C_n^{(p)}|^2 = \frac{D |\T|}{p |\Omega|} \,,
\end{equation*}
where we used the identity (\ref{eq:Vn_int}) and employed the
completeness of the eigenbasis $\{ v_n^{(p)}(\s)\}$.  In
\ref{sec:A_SCA}, we provide some complementary insights and
probabilistic interpretation of the parameters $Q_p$ and $F_p(\x_0)$,
and discuss the short-time and long-time asymptotic behaviors of the
approximate probability density $\rho^{\rm app}(\ell,t)$.

In summary, all three approximations are applicable to multiple
arbitrarily-shaped small targets, which are well separated from each
other and from the outer boundary.  The MMA offers a systematic way to
access higher-order corrections and thus to control the accuracy; the
PEA yields a fully explicit yet accurate expression
(\ref{eq:rho_ell}), which is valid even in higher dimensions; finally,
the SCA remains rather formal due to its need to access spectral
properties of the Dirichlet-to-Neumann operator but yields an exact
solution in some simple domains; it is also less restrictive on the
arrangement of the targets.  While the PEA is probably the most useful
for applications, it is based on the average over a uniformly
distributed starting point and therefore does not capture the impact
of a fixed starting point, which is accessed in both MMA and SCA.
Overall, these three approximations provide complementary tools for
studying the distribution of the boundary local time on small targets.

\section{Comparison for a spherical target}
\label{sec:sphere}

In order to access the accuracy of the approximations that we derived
in Sec. \ref{sec:general}, we consider restricted diffusion toward a
spherical target of radius $R_1$ surrounded by an outer concentric
reflecting sphere of radius $R_2$: $\Omega = \{\x\in\R^3 ~:~ R_1 <
|\x| < R_2\}$.  For this rotation-invariant domain, the eigenmodes of
the Laplace operator and of the Dirichlet-to-Neumann operator are well
known \cite{Grebenkov20c}, in particular,
\begin{equation}  \label{eq:mupI}
\mu_n^{(p)} = - \bigl(\partial_r g_n^{(p)}\bigr)_{|r=R_1} \,,
\end{equation}
with
\begin{equation} \label{eq:gnI}
g_n^{(p)}(r) = \frac{k'_n(\alpha R_2) i_n(\alpha r) - i'_n(\alpha R_2) k_n(\alpha r)}
{k'_n(\alpha R_2) i_n(\alpha R_1) - i'_n(\alpha R_2) k_n(\alpha R_1)}  \,.
\end{equation}
Here $\alpha = \sqrt{p/D}$, $i_n(z) = \sqrt{\pi/(2z)}\, I_{n+1/2}(z)$
and $k_n(z) = \sqrt{2/(\pi z)}\, K_{n+1/2}(z)$ are the modified
spherical Bessel functions of the first and second kind, and the prime
denotes the derivative with respect to the argument.  The radial
functions $g_n^{(p)}(r)$ satisfy the second-order differential
equation
\begin{equation}  \label{eq:gn_diff}
\biggl(\partial_r^2 + \frac{2}{r} \partial_r - \frac{n(n+1)}{r^2} - \alpha^2\biggr) g_n^{(p)}(r) = 0   ,
\end{equation}
with $g_n^{(p)}(R_1) = 1$ and $\bigl(\partial_r g_n^{(p)}\bigr)_{r=R_2} = 0$.

As the target region covers the whole inner sphere, the survival
probability $S_q(t|\x_0)$ and the probability density
$\rho(\ell,t|\x_0)$ of the boundary local time do not depend on the
angular spherical coordinates $(\theta_0,\phi_0)$ of the starting
point $\x_0 = (r_0,\theta_0,\phi_0)$.  In addition, the ground
eigenfunction $v_0^{(p)}(\s) = 1/\sqrt{4\pi R_1^2}$ is constant, and
the projection of other eigenfunctions on $1$ in Eq. (\ref{eq:Cnp})
yields thus
\begin{equation}  \label{eq:Cnp_sphere}
C_n^{(p)} = \sqrt{4\pi} R_1 \, \delta_{n,0}.
\end{equation}
Substituting this expression into Eq. (\ref{eq:rhop_exact}), one has
\begin{equation}
\tilde{\rho}(\ell,p|\x_0) = \tilde{S}_\infty(p|\x_0) \delta(\ell) + \sqrt{4\pi} R_1 \frac{\mu_0^{(p)}}{p} e^{-\mu_0^{(p)}\ell} V_0^{(p)}(\x_0),
\end{equation}
with
\begin{equation}
\tilde{S}_\infty(p|\x_0) = \frac{1 - \tilde{H}_\infty(p|\x_0)}{p} = \frac{1-g_0^{(p)}(r_0)}{p}
\end{equation}
and
\begin{equation}
V_0^{(p)}(\x_0) = \frac{1}{\sqrt{4\pi} R_1} \int\limits_{\T} d\s \, \tilde{j}_\infty(\s,p|\x_0)  
= \frac{\tilde{H}_\infty(p|\x_0)}{\sqrt{4\pi} R_1} = \frac{g_0^{(p)}(r_0)}{\sqrt{4\pi} R_1} \,,
\end{equation}
where $\tilde{H}_\infty(p|\x_0)$ is the Laplace-transformed
probability density of the first-passage time to a perfectly reactive
target.  We get thus
\begin{equation}  \label{eq:rhop_sphere}
\tilde{\rho}(\ell,p|\x_0) = \frac{1-g_0^{(p)}(r_0)}{p} \delta(\ell) + \frac{g_0^{(p)}(r_0)}{p} \mu_0^{(p)} \, e^{-\mu_0^{(p)}\ell} \, .
\end{equation}
When the starting point $\x_0$ is uniformly distributed in $\Omega$,
one can use Eq. (\ref{eq:gn_diff}) to show that
\begin{equation}
\int\limits_{R_1}^{R_2} dr \, r^2 \, g_0^{(p)}(r) = - \frac{R_1^2 (\partial_r g_0^{(p)})_{r =R_1}}{p/D} = \frac{DR_1^2 \mu_0^{(p)}}{p} \,,
\end{equation}
from which
\begin{equation}  \label{eq:rhop_sphere_volume}
\tilde{\rho}(\ell,p) = \biggl(\frac{1}{p} - \frac{4\pi DR_1^2 \mu_0^{(p)}}{p^2 |\Omega|}\biggr)  \delta(\ell) 
+ \frac{4\pi DR_1^2 [\mu_0^{(p)}]^2}{p^2 |\Omega|} e^{-\mu_0^{(p)}\ell} \, .
\end{equation}
Equations (\ref{eq:rhop_sphere}, \ref{eq:rhop_sphere_volume}) provide
the exact form of the probability density in the Laplace domain.
In the following, we focus on Eq. (\ref{eq:rhop_sphere_volume}) and
perform its inverse Laplace transform with respect to $p$ numerically
to get the probability density $\rho(\ell,t)$.  Despite this numerical
step, this solution will be referred to as an exact solution, to which
other approximations will be compared.

The substitution of Eq. (\ref{eq:Cnp_sphere}) into Eqs. (\ref{eq:Qp},
\ref{eq:Fx}) yields
\begin{equation}
Q_p = \mu_0^{(p)} , \qquad F_p(\x_0) = \frac{g_0^{(p)}(r_0)}{\mu_0^{(p)}} \,,
\end{equation}
so that the approximate probability density $\tilde{\rho}^{\rm
app}(\ell,p|\x_0)$ from Eq. (\ref{eq:Pell}) turns out to be identical
with the exact one given by Eq. (\ref{eq:rhop_sphere}).  In other
words, the self-consistent approximation is {\it exact} for the
considered case.  We emphasize that this is a consequence of the
rotation symmetry; in general, the self-consistent approximation does
not coincide with the exact solution.

A comparison with the principal eigenvalue approximation
(\ref{eq:rho_ell}) is straightforward.  In fact, the capacity $C$ of a
spherical target is equal to $4\pi R_1$.  As discussed in
\cite{Chaigneau22}, a more accurate approximation involves the
``corrected'' capacity, which for a spherical target reads as
\begin{equation}  \label{eq:Cprime}
C' = C \bigl(1 - C R(\x_\T,\x_\T)\bigr),
\end{equation}
where $R(\x_T,\x_T)$ is the regular part of the Neumann's Green
function, evaluated at the location $\x_T$ of the target inside the
confining spherical domain of radius $R_2$ \cite{Cheviakov11}:
\begin{align}
(4\pi R_2) \, R(\x,\x) & = \frac{1}{1 - \frac{|\x|^2}{R_2^2}} - \ln \biggl(1 - \frac{|\x|^2}{R_2^2}\biggr) + \frac{|\x|^2}{R_2^2} - \frac{14}{5} \,.
\end{align}
In our case, the target is located at the origin, $\x_T = {\bf 0}$, so
that
\begin{equation}
C' = 4\pi R_1 \biggl(1 + \frac{9R_1}{5R_2} \biggr),
\end{equation}
and thus 
\begin{align}
\lambda_1^{(\infty)} & \approx \frac{C'}{|\Omega|} \approx \frac{3R_1(1 + (9R_1)/(5R_2))}{R_2^3}  \,, \\
L & = \frac{|\T|}{C'} = \frac{R_1}{1 + (9R_1)/(5R_2)} \, ,
\end{align}
where we neglected the small term $R_1^3$ as compared to $R_2^3$ in
the volume $|\Omega|$.

\begin{figure*}
\begin{center}
\includegraphics[width=77mm]{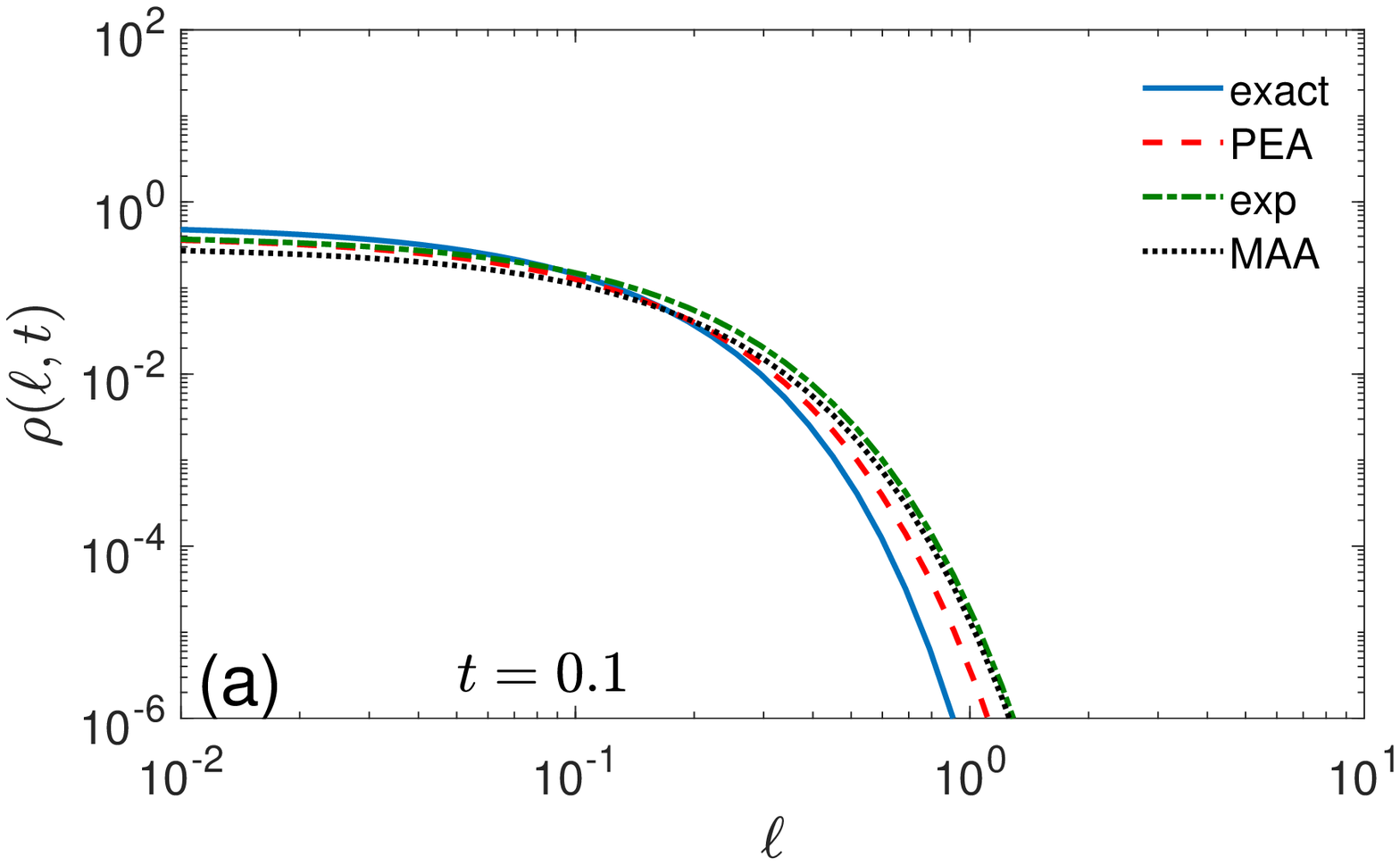} 
\includegraphics[width=77mm]{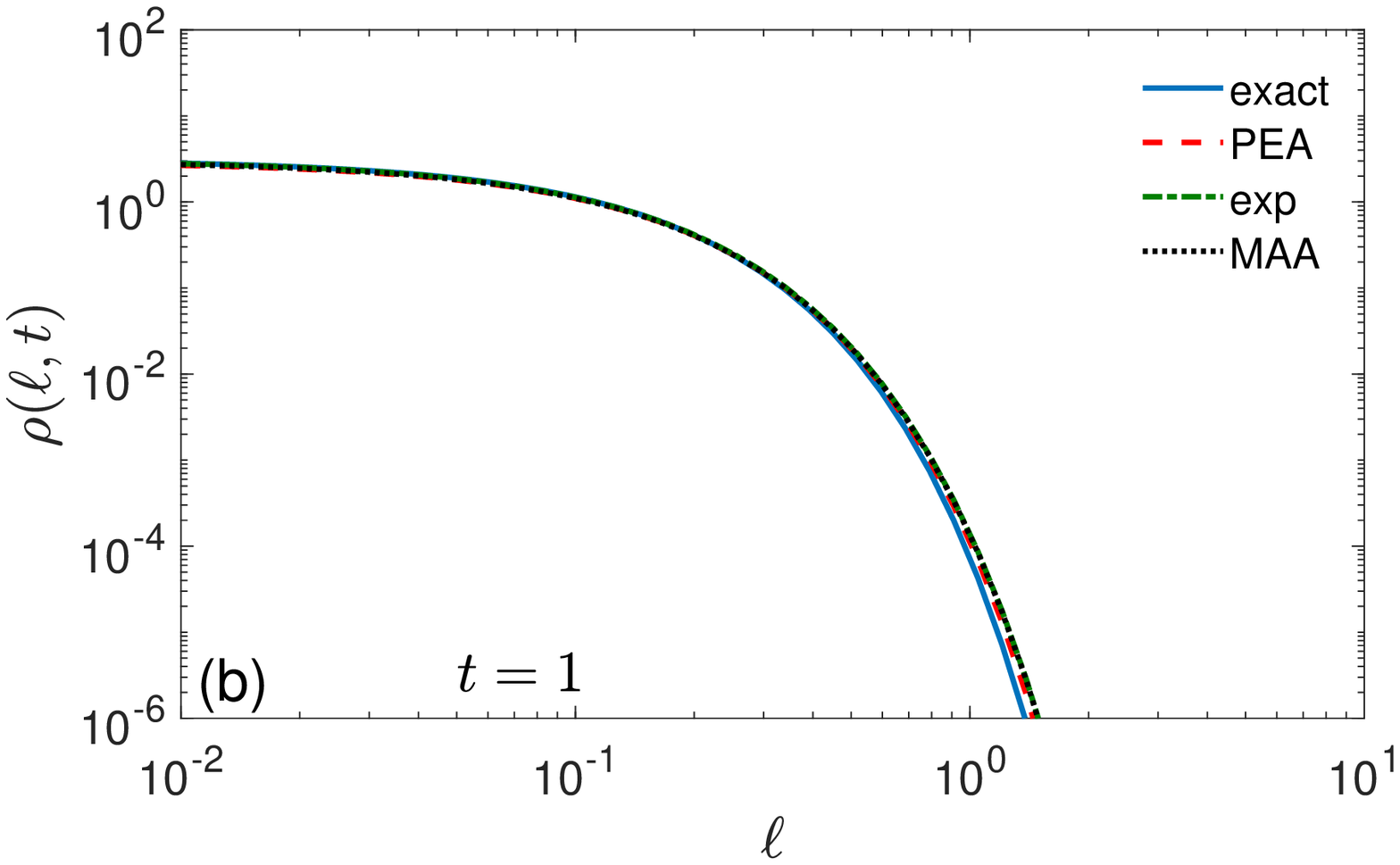} 
\includegraphics[width=77mm]{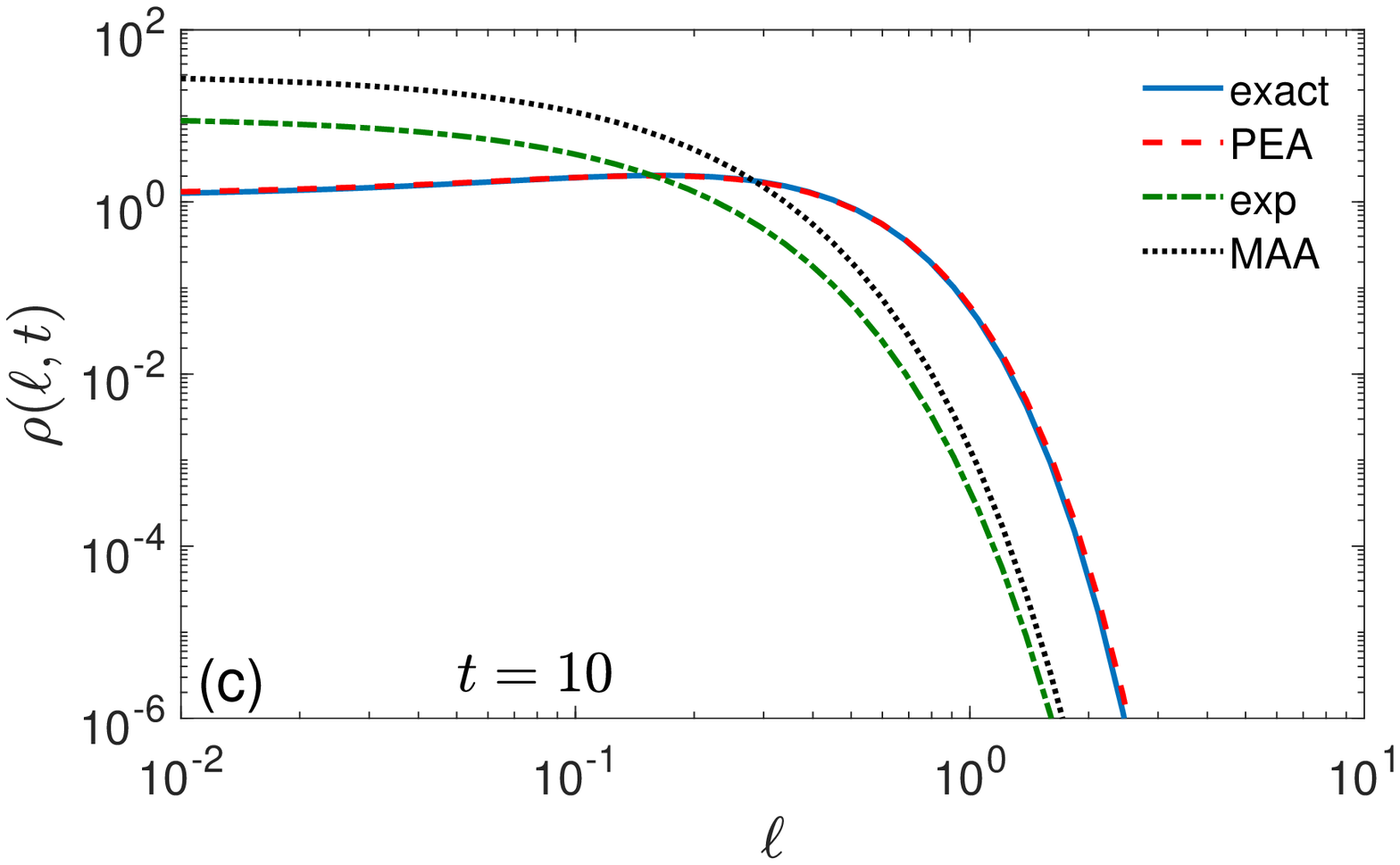} 
\includegraphics[width=77mm]{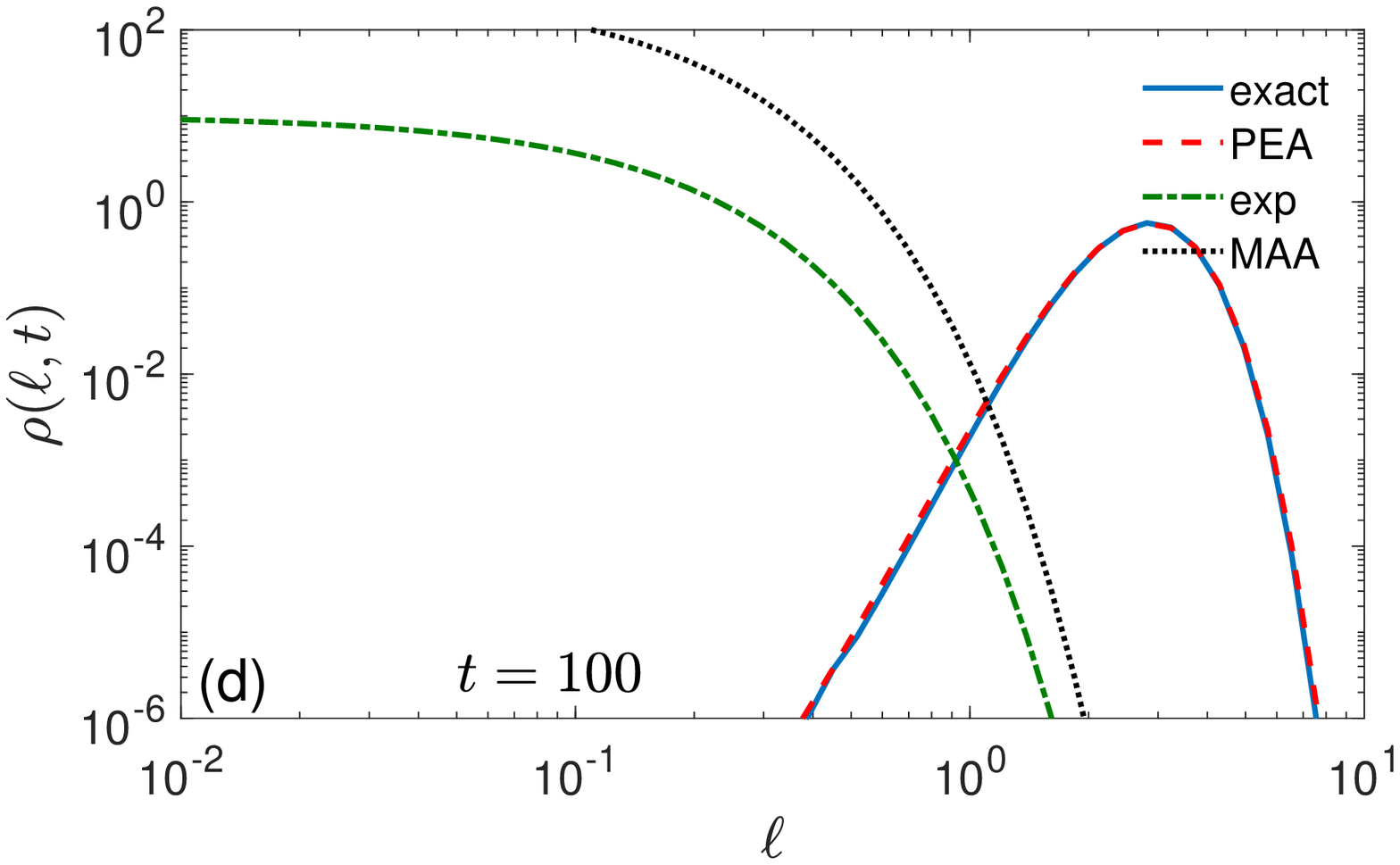} 
\end{center}
\caption{
The regular part of the probability density $\rho(\ell,t)$ of the
boundary local time $\ell_t$ on a spherical target of radius $R_1 =
0.1$ surrounded by a concentric reflecting sphere of radius $R_2 = 1$,
with $D = 1$, the uniform starting point, and several values of time:
$t = 0.1$ {\bf (a)}, $t = 1$ {\bf (b)}, $t = 10$ {\bf (c)}, and $t =
100$ {\bf (d)}.  Solid line presents the benchmark solution obtained
via a numerical inverse Laplace transform of the exact relation
(\ref{eq:rhop_sphere_volume}) by the Talbot algorithm; dashed line
shows the principal eigenvalue approximation (PEA) in
Eq. (\ref{eq:rho_ell}); dash-dotted line shows an exponential
approximation (\ref{eq:rho_exp_av}), in which $S_\infty(t)$ was
computed via its spectral expansion; dotted line indicates the
short-time behavior (\ref{eq:rho_exp_MAA}) predicted by the matched
asymptotic analysis (MMA).}
\label{fig:rho_comparison}
\end{figure*}

Figure \ref{fig:rho_comparison} compares several approximations of the
probability density $\rho(\ell,t)$ to the exact solution obtained via
a numerical inverse Laplace transform of the exact relation
(\ref{eq:rhop_sphere}).  As the singular term $S_\infty(t)
\delta(\ell)$ is the same for all considered approximations, we
present only the regular part of $\rho(\ell,t)$, which allows one to
appreciate the quality of these approximations.
Setting $R_2 = 1$ and $D = 1$ fixes the units of length and time.  In
the considered example, we choose a relatively small target with
$R_1/R_2 = 0.1$, for which the volume-averaged mean first-passage time
is $T \simeq 1/(D\lambda_1^{(\infty)}) \approx R_2^3/(3R_1 D) \approx
3$.  At short times $t$ (as compared to $T$), the survival probability
$S_\infty(t)$ is close to $1$, i.e., most trajectories have not yet
encountered the target, and $\ell_t = 0$.  In this regime, the
singular term $S_\infty(t) \delta(\ell)$ provides the dominant
contribution to $\rho(\ell,t)$.  In turn, the regular part accounts
for contributions from rare trajectories that started from the
vicinity of the target and have encountered it.  The panels {\bf (a)}
and {\bf (b)} of Fig. \ref{fig:rho_comparison} illustrate the
dependence of the regular part on $\ell$ in this regime.  One sees
that three approximations (\ref{eq:rho_exp_MAA}, \ref{eq:rho_exp_av},
\ref{eq:rho_ell}) are in good agreement with the exact solution,
especially in the case $t = 1$.  Some deviations on the panel {\bf
(a)} for $t = 0.1$ will be discussed below.

\begin{figure}
\begin{center}
\includegraphics[width=100mm]{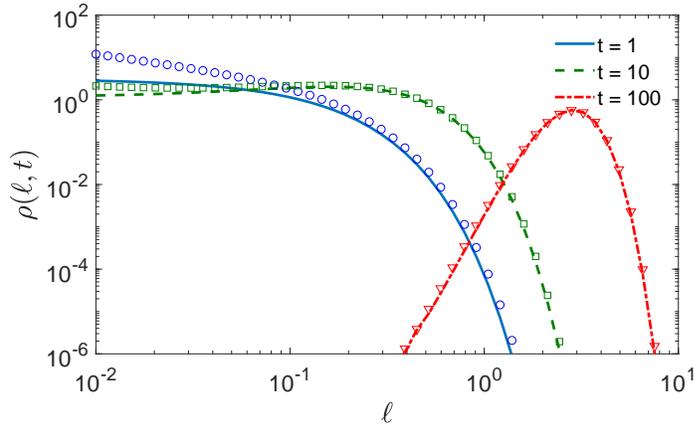} 
\end{center}
\caption{
The regular part of the probability density $\rho(\ell,t)$ of the
boundary local time $\ell_t$ on an inner sphere of radius $R_1 = 0.1$
surrounded by an outer reflecting sphere of radius $R_2 = 1$, with $D
= 1$ and several values of $t$.  Lines present a numerically inverted
Laplace transform of the exact relation (\ref{eq:rhop_sphere_volume})
by the Talbot algorithm, whereas symbols show the large-$\ell$
approximation (\ref{eq:rho_ell_app}).}
\label{fig:long-time}
\end{figure}

In the opposite long-time regime $t \gg T$, $S_\infty(t)$ is close to
$0$, and the dominant contribution comes from the regular part of
$\rho(\ell,t)$ that is shown on panels {\bf (c)} and {\bf (d)} of
Fig. \ref{fig:rho_comparison}.  While the principal eigenvalue
approximation (\ref{eq:rho_ell}) remains to be in excellent agreement
with the exact solution, exponential-like approximations
(\ref{eq:rho_exp_MAA}, \ref{eq:rho_exp_av}) based on the matched
asymptotic analysis, fail at long times.  As discussed in
Sec. \ref{sec:matched}, the leading-order term does not account for
multiple ``long-range returns'' of the particle to the target, which
substantially modify the statistics of the boundary local time.  Even
though the matched asymptotic analysis allows one to obtain next-order
terms, their derivation and resulting expressions rapidly become too
cumbersome for a practical implementation.  In contrast, the principal
eigenvalue approximation (\ref{eq:rho_ell}), which is remarkably
simple and general, fully describes the distribution of $\ell_t$.  Its
excellent quality is rationalized in \ref{sec:A_SCA}, in which
Eq. (\ref{eq:rho_ell}) was re-derived from the long-time asymptotic
behavior of the self-consistent approximation.  We recall that the
self-consistent approximation (\ref{eq:Pell}) itself is identical to
the exact solution for this geometric setting and thus is not
discussed here.
Figure \ref{fig:long-time} illustrates the long-time asymptotic
relation (\ref{eq:rho_ell_app}), which follows from the PEA.
Expectedly, it is accurate at long times $t = 10$ and $t = 100$ but
exhibits deviations at shorter time $t = 1$.

It is instructive to return to the panel {\bf (a)} and inspect
eventual deviations at short times.  Such deviations for the principal
eigenvalue approximation are briefly discussed at the end of
\ref{sec:A_SCA}.  Let us thus focus on deviations of the
exponential-like distribution (\ref{eq:rho_exp_MAA}) predicted by the
matched asymptotic analysis.  As this relation was obtained by using
the short-time approximation, one might expect that
Eq. (\ref{eq:rho_exp_MAA}) would be more and more accurate as $t$ goes
to $0$.  This is not the case.  In fact, the normalization of the
probability density $\rho(\ell,t)$ yields the condition
(\ref{eq:cond_MMA}), which reads as
\begin{equation}  \label{eq:cond_MMA2}
\frac{1-S_\infty(t)}{R} \approx \frac{4\pi Dt}{|\Omega|} \,.
\end{equation}
This relation can be considered as a necessary condition for the
consistence of the approximation (\ref{eq:rho_exp_MAA}).  Clearly,
this relation fails at long times, at which $S_\infty(t)$ vanishes.
Is it correct at short times?  The answer is negative.  In fact, the
short-time asymptotic behavior of the survival probability follows
from the heat content asymptotics
\cite{vandenBerg89,vandenBerg94,Desjardins94,Gilkey} and reads
\cite{Grebenkov20e}
\begin{equation}  \label{eq:Sinfty_t0}
S_\infty(t) \approx 1 - \frac{2|\T|}{\sqrt{\pi} |\Omega|} \sqrt{Dt} + O(t).
\end{equation}
In other words, since the left-hand side of Eq. (\ref{eq:cond_MMA2})
scales as $t^{1/2}$, while the right-hand side does as $t$, this
relation cannot hold at short times.  Fortunately, there is an
intermediate range of time scales at which Eq. (\ref{eq:cond_MMA2})
holds.  In fact, if the target is small enough, the survival
probability can be approximated via Eq. (\ref{eq:Sq_approx}), which at
moderately short times admits the Taylor expansion:
\begin{equation}
S_\infty(t) \approx 1 - Dt\lambda_1^{(\infty)} + O(t^2) .
\end{equation}
and thus Eq. (\ref{eq:cond_MMA}) is equivalent to
$\lambda_1^{(\infty)} \approx 4\pi R_1/|\Omega|$.  As the numerator of
the right-hand side is precisely the capacity of a spherical target of
radius $R_1$, we retrieve the approximation given in
Eq. (\ref{eq:lambda1}).  In other words, even though the condition
(\ref{eq:cond_MMA2}) fails in both limits $t\to 0$ and $t\to\infty$, it
is fulfilled at intermediate times if the target is small enough.  We
conclude that the exponential-like distribution (\ref{eq:rho_exp_MAA})
is valid whenever the principal eigenvalue approximation
(\ref{eq:rho_ell}) is.  The latter is therefore much more general.

To complete this discussion, we provide some qualitative arguments why
the predictions of the matched asymptotic analysis fail at very short
times.  In this regime, the only nontrivial contribution comes from a
very thin layer near the target.  At the same time, the main idea of
the matched asymptotic analysis consists in matching inner and outer
solutions of the problem.  For the other solution, the target is
treated as point-like (for instance, Eq. (\ref{eq:tildeU0}) involves
the fundamental solution $\tilde{g}(\x,p|\x_0)$ of the modified
Helmholtz equation, which deals with a point-like source), so that its
geometric structure is not accessible at short times.  In other words,
the matched asymptotic analysis focuses on the limit, in which the
target size is the smallest parameter of the problem; in particular,
$R_1$ should be much smaller than a diffusion length $\sqrt{Dt}$.
This condition yields a natural restriction of considering not too
short times, namely, $t \gg R_1^2/D$.  This qualitative argument
explains why the agreement on panel {\bf (a)} for $t = 0.1$ is still
acceptable, despite some minor deviations.  In turn, deviations are
more significant for even small $t$ (not shown).

\section{Distribution of first-crossing times}
\label{sec:discussion}

As the principal eigenvalue approximation turns out to be the most
convenient, one can apply it to investigate other properties of the
boundary local time and related quantities.  
Integrating the series expansion (\ref{eq:rho_series}) of $\rho_{\rm
PEA}(\ell,t)$ term by term, we also get an approximation
\begin{align}    \nonumber
\P\{\ell_t > \ell\} & = \int\limits_\ell^\infty d\ell' \, \rho(\ell',t) 
\approx \int\limits_\ell^\infty d\ell' \, \rho_{\rm PEA}(\ell',t) \\  \label{eq:Pell_series}
& = e^{-t/T - \ell/L} \sum\limits_{n=0}^\infty \frac{(t/T)^{n+1}}{(n+1)!}
\sum\limits_{k=0}^n \frac{(\ell/L)^k}{k!} 
\end{align}
that determines the cumulative distribution function of $\ell_t$.
Alternatively, one can integrate directly the final expression
(\ref{eq:rho_ell}), which yields after changing the integration
variable
\begin{equation}  
\P\{\ell_t > \ell\} \approx 2 \sqrt{t/T} \, e^{- t/T} \int\limits_{\sqrt{\ell/L}}^\infty dz \, e^{-z^2} \, 
I_1\bigl(2z \sqrt{t/T}\bigr).
\end{equation}

The function $\P\{\ell_t > \ell\}$ determines also the distribution of
the first-crossing time $\TT_\ell = \inf\{ t>0~:~ \ell_t > \ell\}$ of
a given threshold $\ell$ by the boundary local time $\ell_t$, which
plays an important role in diffusion-controlled reactions
\cite{Grebenkov20a}.  In fact, since $\ell_t$ is a non-decreasing
process, one has $\P\{\ell_t > \ell\} = \P\{\TT_\ell < t\}$, from
which the probability density of $\TT_\ell$ follows as
\begin{equation}
U(\ell,t) = \partial_t \P\{\TT_\ell < t\} = \partial_t \P\{\ell_t > \ell\} .
\end{equation}
Evaluating the time derivative of Eq. (\ref{eq:Pell_series}) term by
term, we get the following approximation:
\begin{equation}  \label{eq:Ut_PEA}
U_{\rm PEA}(\ell,t) = \frac{1}{T}\, e^{-\ell/L - t/T} \, I_0\biggl(2\sqrt{(t/T)(\ell/L)}\biggr) \,.
\end{equation}
This is one of the main results of the paper.  Despite its approximate
character, this probability density is correctly normalized.  For
large $\ell$ or $t$, the asymptotic behavior of $I_0(z)$ yields
\begin{equation}  \label{eq:Ut_long2}
U_{\rm PEA}(\ell,t) \approx \frac{\exp\biggl(-\biggl(\sqrt{\ell/L} - \sqrt{t/T}\biggr)^2\biggr)}
{2\sqrt{\pi} (T)^{\frac34} \, t^{\frac14} \, (\ell/L)^{\frac14}} \,.
\end{equation}
The moments of $\TT_\ell$ can be easily accessed:
\begin{align}  \nonumber
\E\{ [\TT_\ell]^k\} & = \int\limits_0^\infty dt \, t^k \, U(\ell,t) \approx T^k \, e^{-\ell/L}
\int\limits_0^\infty dz \, z^k \, e^{-z} \, I_0\bigl(2\sqrt{z\ell/L}\bigr) \\
& = T^k \, k! \, e^{-\ell/L} M(k+1;1; \ell/L).
\end{align}
For instance, one gets
\begin{equation}
\E\{ \TT_\ell \} \approx (1 + \ell/L)T \,, \qquad
\mathrm{var}\{ \TT_\ell\} \approx (1 + 2\ell/L) T^2 \,.
\end{equation}

\begin{figure}
\begin{center}
\includegraphics[width=100mm]{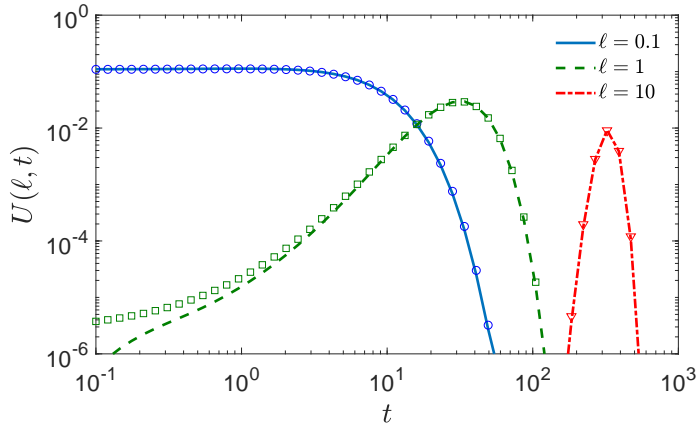} 
\end{center}
\caption{
The probability density $U(\ell,t)$ of the first-crossing time
$\TT_\ell$ of the boundary local time $\ell_t$ on an inner sphere of
radius $R_1 = 0.1$ surrounded by an outer reflecting sphere of radius
$R_2 = 1$, with $D = 1$ and several values of the threshold $\ell$.
Lines present a numerically inverted Laplace transform of the exact
relation (\ref{eq:Ut_sphere}) by the Talbot algorithm, whereas symbols
show the principal eigenvalue approximation (\ref{eq:Ut_PEA}).}
\label{fig:Ut}
\end{figure}

Note that the probability density $U(\ell,t)$ can formally be accessed
from the general exact representation (\ref{eq:rho_exact_volume}),
which yields for the spherical case:
\begin{equation}  \label{eq:Ut_sphere}
U(\ell,t) = \L^{-1}_p\biggl\{\frac{4\pi D R_1^2 \mu_0^{(p)}}{p |\Omega|} e^{-\mu_0^{(p)}\ell}  \biggr\} .
\end{equation}
A numerical inversion of this Laplace transform can be used as a
benchmark for validating our approximation.  Figure \ref{fig:Ut}
illustrates an excellent accuracy of the approximation
(\ref{eq:Ut_PEA}) for several values of the threshold $\ell$ when $t$
is not too small.  In turn, the approximation fails in the small-time
limit.  In fact, Eq. (\ref{eq:Ut_PEA}) suggests that $U(\ell,t) \to
D\lambda_1^{(\infty)} e^{-\ell/L}$ as $t\to 0$, whereas the exact
solution (\ref{eq:Ut_sphere}) implies a rapid decay of $U(\ell,t)$.
This can be seen from the large-$p$ asymptotic analysis of
Eq. (\ref{eq:mupI}) that yields $\mu_0^{(p)} \approx \sqrt{p/D} +
1/R_1$ for the considered spherical case, from which the inverse
Laplace transform in Eq. (\ref{eq:Ut_sphere}) implies
\begin{equation}
U(\ell,t) \approx \frac{|\T| D}{|\Omega|} e^{-\ell/R_1} \biggl(\frac{e^{-\ell^2/(4Dt)}}{\sqrt{\pi Dt}} 
+ \frac{1}{R_1} \erfc\biggl(\frac{\ell}{\sqrt{4Dt}}\biggr) \biggr)  \quad (t\to 0).
\end{equation}
The short-time deviation of the approximation (\ref{eq:Ut_PEA}) is
clearly seen in Fig. \ref{fig:Ut} for the case $\ell = 1$.  Despite
this limitation at short times, the simple explicit form of
Eq. (\ref{eq:Ut_PEA}) captures the behavior of the probability density
$U(\ell,t)$ remarkably well.

\section{Conclusion}
\label{sec:conclusion}

We studied the distribution of the boundary local time $\ell_t$, i.e.,
a rescaled number of encounters between a diffusing particle and a
target.  This distribution can formally be obtained either by the
inverse Laplace transform of the survival probability with respect to
the reactivity parameter $q$, or as an expansion over the eigenbasis
of the Dirichlet-to-Neumann operator.  In both cases, the underlying
quantities depend on the shapes of the confining domain and of the
target in a sophisticated, generally unknown way.  To access this
distribution in the case of a {\it small} target, we employed three
approximations: the matched asymptotic analysis, the principal
eigenvalue approximation, and the self-consistent approximation.  The
leading order of the MAA yielded an exponential-like distribution
(\ref{eq:rho_exp_av}), with an atom at $\ell = 0$ corresponding to the
trajectories that never encountered the target.  This approximation
was shown to be accurate only at intermediate times but failing in
both short- and long-time limits.  Such an approximation can be
potentially improved by considering higher-order corrections in the
MAA which, however, are much more cumbersome.  In turn, the principal
eigenvalue approximation provided a simple, fully explicit and
remarkably accurate approximation (\ref{eq:rho_ell}).  This
approximation involves only few geometric characteristics such as the
surface area and the harmonic capacity of the target, as well as the
volume of the confining domain.  In particular, it reveals how the
boundary local time $\ell$ is coupled to physical time $t$.  This is a
rare example of an explicitly known distribution of the boundary local
time (for two other basic examples, the half-line and the exterior of
a sphere, see \cite{Grebenkov21a}).  The third approach was based on
the self-consistent approximation, which yielded a compact form of the
probability density of the boundary local time in the Laplace domain.
The need for the Laplace inversion to come back in time domain makes
this approximation less appealing in comparison to the PAE.
Nevertheless, the SCA-based relations (\ref{eq:Pell},
\ref{eq:rho_app0}) allowed us to study the asymptotic behaviors; in
particular, we managed to retrieve the PEA (\ref{eq:rho_ell}) as the
long-time asymptotic limit of the SCA.  The accuracy of different
approximations was checked in a typical geometric setting of a
spherical target surrounded by a reflecting sphere.

It is worth noting that the inversion of the Laplace transform
is actually not needed for some applications.  For instance,
when a particle diffuses in a reactive medium or has an internal
destructive mechanism (aging, radioactive decay, photobleaching,
nuclear spin relaxation, etc.), its lifetime $\tau$ is random and
usually described by an exponential law with the decay rate $p$
\cite{Yuste13,Meerson15,Grebenkov17c}.   In this case, the above
approximations give a direct access to the probability density $p \,
\tilde{\rho}(\ell,p|\x_0)$ of the boundary local time $\ell_\tau$,
which is stopped at a random time $\tau$.  Here, $\ell_\tau$
characterizes the number of encounters with the target until the
particle's death.  For instance, Eq. (\ref{eq:Pell}) shows that $p\,
\tilde{\rho}(\ell,p|\x_0)$ admits a simple exponential-like form, with
an atom at $0$.  Moreover, the small-$p$ expansion of the
Laplace-transformed probability density $\tilde{\rho}(\ell,p|\x_0)$
determines the moments of the boundary local time $\ell_t$.  Other
quantities such as splitting probabilities and conditional
first-passage time moments can also be accessed \cite{Bressloff22}.

This paper was mainly focused on three-dimensional confining domains.
In particular, this choice allowed us to compare three approximations
given that the matched asymptotic analysis was done by Bressloff only
in three dimensions \cite{Bressloff22}.  Its extension to other space
dimensions is in principle feasible but requires additional work.  In
turn, the self-consistent approximation was formulated in terms of the
eigenbasis of the Dirichlet-to-Neumann operator and is thus valid for
any dimension.  Finally, the principal eigenvalue approximation
(\ref{eq:lambda1}) is applicable for any $d \geq 3$; moreover, it is
getting more accurate as $d$ increases \cite{Chaigneau22}.  We
therefore expect that our approximation (\ref{eq:rho_ell}) for the
probability density $\rho(\ell,t)$ would also be accurate in higher
dimensions.  In turn, the two-dimensional case was excluded from the
analysis in \cite{Chaigneau22}.  In fact, the harmonic capacity does
not exist in two dimensions, whereas solutions of the Laplace equation
are much more sensitive to a distant outer boundary.  At the same
time, one can still consider Eq. (\ref{eq:lambda1}) as an
interpolation between two limits of perfectly reactive ($q\to \infty$)
and weakly reactive ($q\to 0$) targets.  Setting $C =
\lambda_1^{(\infty)} |\Omega|$ and thus $L =
|\T|/(\lambda_1^{(\infty)} |\Omega|)$, one sees that
Eq. (\ref{eq:lambda1}) interpolates between the limit
$\lambda_1^{(\infty)}$ as $q\to \infty$ and the expected behavior
$\lambda_1^{(q)} \approx q |\T|/|\Omega|$ as $q\to 0$.  Admitting this
interpolation, one can keep the derivation in
Sec. \ref{sec:eigenvalue} and thus retrieve again the approximation
(\ref{eq:rho_ell}) in two dimensions.  Moreover, the eigenvalue
$\lambda_1^{(\infty)}$ can also be related to the geometric properties
of a small target \cite{Mazya85,Ward93,Kolokolnikov05}.  Numerical
validation of this conjectural extension to the two-dimensional case
presents an interesting perspective.

While we were mainly interested in the distribution of the boundary
local time $\ell_t$, the approximation (\ref{eq:rho_ell}) allowed us
to access the tightly related probability density $U(\ell,t)$ of the
first-crossing time $\TT_\ell$ of a given threshold $\ell$ by
$\ell_t$.  Our approximation (\ref{eq:Ut_PEA}) is a rare example when
this probability density admits a fully explicit form.
As discussed in \cite{Grebenkov20a}, this distribution is directly
related to the distribution of first-reaction times on a partially
reactive target.  Moreover, one can go beyond the conventional
constant reactivity framework (described by the Robin boundary
condition) and deal with other surface reaction mechanisms such as
encounter-dependent reactivity.  A simple explicit form of the
probability density $U(\ell,t)$ allows one to investigate a broad
class of first-passage times related to various surface reaction
mechanisms.  The explicit form of $U(\ell,t)$ can also be employed to
study in more detail the resource depletion problem by a population of
diffusing particles \cite{Grebenkov22f}.

\appendix

\section{Analysis of the self-consistent approximation}
\label{sec:A_SCA}

In this Appendix, we discuss the probabilistic interpretation of the
parameters $Q_p$ and $F_p(\x_0)$ that determine the self-consistent
approximation, as well as its short-time and long-time asymptotic
behaviors.

\subsection{Complementary insights}

Let us gain complementary insights onto the parameters $Q_p$ and
$F_p(\x_0)$.  Since the kernel of the Dirichlet-to-Neumann operator
$\M_p$ is $D \tilde{G}_0(\s,p|\s_0)$ (see
\cite{Grebenkov20a}), the expression (\ref{eq:Qp}) can also be written
as
\begin{equation}  \label{eq:Qp2}
\frac{1}{Q_p} = \frac{1}{|\T|} \int\limits_{\T} d\s_0 \int\limits_{\T} d\s \, D \tilde{G}_0(\s,p|\s_0).
\end{equation}
In order to interpret this expression, we recall that the mean
residence time in a subset $A\subset \Omega$ is
\begin{align*}
R_{A}(t|\x_0) &= \E_{\x_0}\left\{ \int\limits_0^t dt' \I_{A}(\X_{t'}) \right\}  
 = \int\limits_0^t dt' \int\limits_{A} d\x  \, \E_{\x_0}\{\delta(\x - \X_{t'}) \} \\
& = \int\limits_0^t dt' \int\limits_{A} d\x \, G_0(\x,t'|\x_0) .
\end{align*}
If $A$ is a thin layer near the target, $A = \T_a$, then we get the
mean boundary local time on $\T$:
\begin{equation}
\E_{\x_0}\{ \ell_t\} = \lim\limits_{a\to 0} \frac{D}{a} R_{\T_a}(t|\x_0) = \int\limits_0^t dt' \int\limits_{\T} d\x \, D G_0(\x,t'|\x_0),
\end{equation}
where we used that $G_0(\x,t'|\x_0)$ behaves smoothly in a vicinity of
a smooth boundary.  If the particle has a finite lifetime, one has to
average over $t$ with an exponential probability density:
\begin{equation}
\E_{\x_0}\{\ell_\tau\} = \int\limits_0^\infty dt \, \underbrace{p e^{-pt}}_{\textrm{pdf of}~\tau}\, 
\E_{\x_0}\{ \ell_t\} = \int\limits_{\T} d\x \, D \tilde{G}_0(\x,p|\x_0).
\end{equation}
Note that the inverse Laplace transform of $\E_{\x_0}\{\ell_\tau\}/p$
yields the mean $\E_{\x_0}\{\ell_t\}$.  We conclude that $1/Q_p$ in
Eq. (\ref{eq:Qp2}) is the mean boundary local time $\ell_\tau$ on
$\T$, averaged over the starting point $\x_0$ uniformly distributed on
$\T$.

For a fixed starting point, one can use Eq. (\ref{eq:Pell}) to
approximate the moments of the boundary local time $\ell_\tau$, in
particular, the mean is
\begin{equation}
\E_{\x_0}\{ \ell_\tau \} \approx \int\limits_0^\infty d\ell \, \ell \, p\, \tilde{\rho}^{\rm app}(\ell,p|\x_0) = F_p(\x_0) \,.
\end{equation}
Its average over $\x_0 \in \T$ reads
\begin{equation}  \label{eq:average_Fp}
\frac{1}{|\T|} \int\limits_{\T} d\x_0 \, F_p(\x_0) = \frac{1}{Q_p} \,,
\end{equation}
so that
\begin{equation}
\frac{1}{|\T|} \int\limits_{\T} d\x_0 \, p\, \tilde{\rho}^{\rm app}(\ell,p|\x_0) = Q_p \,\exp(-Q_p \ell)   .
\end{equation}
In summary, if the starting point is uniformly distributed over the
target $\T$, the distribution of the boundary local time $\ell_\tau$
is close to be exponential, with the mean value $1/Q_p$.

\subsection{Long-time behavior}

It is instructive to look at the long-time behavior of the approximate
probability density $\rho^{\rm app}(\ell,t)$ that follows from the
small-$p$ asymptotic analysis of Eq. (\ref{eq:rho_app0}).

In the limit $p\to 0$, one can apply a standard perturbation theory
discussed in \cite{Grebenkov19b}.  Adapting this approach to our
setting with a target $\T$ surrounded by a reflecting boundary
$\pa_0$, we get
\begin{equation}
\mu_0^{(p)} \approx \frac{|\Omega|}{D|\T|} p + \frac{bp^2}{2} + O(p^3),  
\end{equation}
where the parameter $b$ can be represented as
\begin{align*}
b & \equiv \lim\limits_{p\to 0} \frac{d^2 \mu_0^{(p)}}{dp^2} 
= - \frac{2|\Omega|^2}{D|\T|^3} \int\limits_{\T} d\s_1 \int\limits_{\T} d\s_2 \, {\mathcal G}(\s_1,\s_2) ,
\end{align*}
with
\begin{equation}
{\mathcal G}(\s_1,\s_2) = \lim\limits_{p\to 0} \biggl(\tilde{G}_0(\s,p|\s_0) - \frac{1}{p|\Omega|}\biggr)
\end{equation}
being the Neumann pseudo-Green's function.  In turn, the corresponding
eigenfunction behaves as $v_0^{(p)}(\s) \approx |\T|^{-1/2} + p
v_0^{1}(\s) + O(p^2)$ as $p \to 0$.  As a consequence, $C_n^{(p)}
\approx O(p)$ for $n > 0$ due to the orthogonality of eigenfunctions
$v_n^{(p)}$ to $v_0^{(p)}$ (and thus to a constant), in the leading
order in $p$.  In turn, the normalization of the eigenfunction
$v_0^{(p)}$ implies
\begin{equation*}
1 = \|v_0^{(p)}\|^2_{L_2(\T)} = \underbrace{\|v_0^{(0)}\|^2_{L_2(\T)}}_{=1} + 2p\bigl(v_0^{(0)} \cdot v_0^1\bigr)_{L_2(\T)} + O(p^2),
\end{equation*} 
and thus $(v_0^{(0)} \cdot v_0^1)_{L_2(\T)} = 0$, i.e., the integral
of the first-order correction $v_0^1(\s)$ vanishes.  As a consequence,
$C_0^{(p)} \approx |\T|^{1/2} + O(p^2)$ (i.e., there is no $O(p)$
term).  Substituting these relations into Eq. (\ref{eq:Qp}), we get
\begin{equation}  \label{eq:Qp0}
Q_p^{-1} \approx \frac{D|\T|}{p|\Omega|} + L + O(p),  \qquad  \textrm{with} \quad L = -\frac{b}{2} (D|\T|/|\Omega|)^2 ,
\end{equation}
which can also be written as
\begin{equation}
Q_p \approx \frac{|\Omega|}{D|\T|} \, \frac{p}{1 + p\ttau}  \,, \qquad \textrm{with} \quad  \ttau = L \frac{|\Omega|}{D|\T|} \,,
\end{equation}
where we neglected the next-order terms.  Substituting this expression
into Eq. (\ref{eq:rho_app0}), we get
\begin{align}
\tilde{\rho}^{\rm app}(\ell,p) & \approx \frac{\ttau}{1 + p\ttau} \delta(\ell) + \frac{\ttau/L}{(1+p\ttau)^2} \, e^{- (\ell/L) p\ttau/(1 + p\ttau)} \,.
\end{align}
The inverse Laplace transform with respect to $p$ reads
\begin{align}  \nonumber
\rho^{\rm app}(\ell,t) & \approx e^{-t/\ttau} \delta(\ell) + \frac{\ttau}{L} e^{-\ell/L}  \L^{-1}_p\biggl\{
\frac{e^{(\ell/L)/(1 + p\ttau)}}{(1+p\ttau)^2} \biggr\}  \\  \nonumber
& = e^{-t/\ttau} \delta(\ell) + \frac{\ttau}{L} e^{-\ell/L} \sum\limits_{n=0}^\infty \frac{(\ell/L)^n}{n!} 
\L^{-1}_p\biggl\{\frac{1}{(1+p\ttau)^{n+2}} \biggr\} \\  \nonumber
& = e^{-t/\ttau} \delta(\ell) + e^{-\ell/L} e^{-t/\ttau} \frac{\sqrt{t/\ttau}}{\sqrt{\ell L}} 
\sum\limits_{n=0}^\infty \frac{((\ell/L)t/\ttau)^{n+\frac12}}{n!(n+1)!} \\   \label{eq:Papp}
& = e^{-t/\ttau} \delta(\ell) + e^{-\ell/L} e^{-t/\ttau} \frac{\sqrt{t/\ttau}}{\sqrt{\ell L}} I_1\bigl(2\sqrt{(\ell/L)t/\ttau}\bigr) ,
\end{align}

Remarkably, the long-time asymptotic relation (\ref{eq:Papp})
coincides with the principal eigenvalue approximation
(\ref{eq:rho_ell}) from Sec. \ref{sec:eigenvalue}, with a different
notation $L = |\T|/C$, where $C$ is the capacity of the target.  Using
this relation, we obtain an interesting representation for the second
derivative of $\mu_0^{(p)}$:
\begin{equation}
b \approx - \frac{2|\Omega|^2}{C |\T| D^2} \,.
\end{equation}
Moreover, substituting this expression into the definition of $\ttau$,
we get $\ttau = |\Omega|/(DC) = 1/(D\lambda_1^{(\infty)})$, i.e., we
retrieve an approximation $\lambda_1^{(\infty)} = C/|\Omega|$ for the
principal eigenvalue for the perfect target.  In this way, two
approaches result in the same long-time behavior.

\subsection{Short-time behavior}

The short-time behavior corresponds to the large-$p$ limit.  As the
boundary region $\T$ is smooth, the propagator $G_0(\x,t|\x_0)$ is
close to that in the half-space with reflecting hyperplane:
\begin{align*}
G_0^{\rm half}(\x,t|\x_0) & = \frac{e^{-|\s-\s_0|^2/(4Dt)}}{(4\pi Dt)^{(d-1)/2}} \, 
\frac{e^{-(y-y_0)^2/(4Dt)} + e^{-(y+y_0)^2/(4Dt)}}{\sqrt{4\pi Dt}} \,,
\end{align*}
where $\x = (\s,y)$ and $\x_0 = (\s_0,y_0)$.  Its Laplace transform
reads
\begin{align*}
\tilde{G}_0^{\rm half}(\x,p|\x_0) & = \frac{(D/p)^{\nu/2}}{(2\pi)^{d/2} D} \biggl( A_1^\nu K_\nu \bigl(A_1\sqrt{p/D}\bigr) 
 + A_2^\nu K_\nu\bigl(A_2\sqrt{p/D}\bigr)\biggr),
\end{align*}
where $\nu = 1 - d/2$, $A_1^2 = |\s-\s_0|^2 + (y-y_0)^2$, $A_2^2 =
|\s-\s_0|^2 + (y+y_0)^2$, and $K_\nu(z)$ is the modified Bessel
function of the second kind.  Setting $y = y_0 = 0$, one gets
\begin{equation}
\tilde{G}_0^{\rm half}(\s,p|\s_0) = \frac{2(D/p)^{\nu/2}}{(2\pi)^{d/2} D}\, |\s-\s_0|^\nu K_\nu\bigl(|\s-\s_0|\sqrt{p/D}\bigr) .
\end{equation}
At large $p$, Eq. (\ref{eq:Qp}) implies thus
\begin{align*}
\frac{1}{Q_p} & \approx \frac{1}{|\T|} \int\limits_{\T} d\s_0 \int\limits_{\T} d\s \, D \tilde{G}_0^{\rm half}(\s,p|\s_0) \\
& \approx \frac{2(D/p)^{\nu/2}}{(2\pi)^{d/2} |\T|} \int\limits_{\T} d\s_0 \int\limits_{\T} d\s \, 
|\s-\s_0|^\nu K_\nu\bigl(|\s-\s_0|\sqrt{p/D}\bigr).
\end{align*}
As $K_\nu(z)$ decays exponentially at large $z$, the main contribution
comes from the points $\s \approx \s_0$.  One integral yields thus
$|\T|$.  In turn, the second integral can be evaluated by replacing
the target surface by a hyperplane and using the spherical
coordinates:
\begin{align*}
\frac{1}{Q_p} & \approx \frac{2(D/p)^{\nu/2}}{(2\pi)^{d/2}} \sigma_{d-1} \int\limits_0^\infty dr\, r^{d-2} \, r^\nu K_\nu\bigl(r\sqrt{p/D}\bigr).
\end{align*}
where $\sigma_{d-1} = 2\pi^{(d-1)/2}/\Gamma((d-1)/2)$ is the surface
area of the unit ball in $\R^{d-1}$, and we extended the upper limit
of integration to infinity.  The last integral can be found exactly
via the identity:
\begin{equation*}
\int\limits_0^\infty dz \, z^\alpha \, K_\nu(z) = 2^{\alpha-1} \Gamma\biggl(\frac{\alpha+\nu+1}{2}\biggr) \Gamma\biggl(\frac{\alpha-\nu+1}{2}\biggr).
\end{equation*}
After simplifications, we get simply
\begin{equation}  \label{eq:Qp_small}
Q_p \approx \sqrt{p/D}  \qquad (p\to \infty).
\end{equation}
Substituting this expression into Eq. (\ref{eq:rho_app0}), we can
invert the Laplace transform to get the short-time approximation:
\begin{equation}  \label{eq:rho_t0}
\rho^{\rm app}(\ell,t) \approx \biggl(1 - \frac{2\sqrt{Dt}\, |\T|}{\sqrt{\pi} |\Omega|}\biggr) \delta(\ell)
+ \frac{|\T|}{|\Omega|} \erfc\biggl(\frac{\ell}{\sqrt{4Dt}}\biggr) \qquad (t\to 0).
\end{equation}
Despite its approximation character, this expression is correctly
normalized.

We emphasize that Eq. (\ref{eq:Qp_small}) and the consequent relation
(\ref{eq:rho_t0}) ignore the curvature of the target, which is known
to yield non-universal corrections \cite{Grebenkov22f}.  For instance,
if the target is a sphere of radius $R_1$, Eq. (\ref{eq:mupI}) implies
$\mu_0^{(p)} \approx \sqrt{p/D} + 1/R_1$, and its substitution into
Eq. (\ref{eq:rhop_sphere_volume}) and the evaluation of the inverse
Laplace transform yield the following short-time asymptotic behavior:
\begin{align}
& \rho(\ell,t) \approx \biggl\{ 1 - \frac{|\T|}{|\Omega|} \biggl(\frac{2\sqrt{Dt}}{\sqrt{\pi}} + \frac{Dt}{R_1}\biggr)\biggr\} \delta(\ell) \\ \nonumber
& + \frac{|\T|}{|\Omega|} e^{-\ell/R_1} \biggl\{ \biggl(1 - \frac{2\ell}{R_1} + \frac{\ell^2 + 2Dt}{2R_1^2}\biggr)
\erfc\biggl(\frac{\ell}{\sqrt{4Dt}}\biggr) + \frac{(4 - \ell/R_1)  \sqrt{Dt}}{R_1 \sqrt{\pi}} e^{-\ell^2/(4Dt)}\biggr\} .
\end{align}
Its comparison with Eq. (\ref{eq:rho_t0}) reveals that the regular
part of this expression contains in an additional factor
$e^{-\ell/R_1}$ due to the curvature, as well as higher-order
corrections.  The above expression is valid for $\sqrt{Dt} \ll R_1$.
One can distinguish two regimes depending on whether $\ell$ is smaller
or larger than $\sqrt{Dt}$. When $\sqrt{Dt} \ll \ell$, one gets in the
leading-order:
\begin{equation}
\rho(\ell,t) \approx \biggl( 1 -  \frac{2\sqrt{Dt} \,|\T|}{\sqrt{\pi} |\Omega|} \biggr) \delta(\ell) 
+ \frac{2\sqrt{Dt}\, |\T|}{\sqrt{\pi} \,|\Omega| \, \ell} e^{-\ell^2/(4Dt)} e^{-\ell/R_1}  \,.
\end{equation}
In the opposite regime $\ell \ll \sqrt{Dt}$, the probability density
approaches 
\begin{equation}
\rho(\ell,t) \approx \biggl( 1 -  \frac{2\sqrt{Dt} \,|\T|}{\sqrt{\pi} |\Omega|} \biggr) \delta(\ell) 
+ \frac{|\T|}{|\Omega|} \biggl(1 + \frac{4\sqrt{Dt}}{\sqrt{\pi} R_1} + \frac{Dt}{R_1^2} \biggr)  \,.
\end{equation}
This relation shows the limitation of the principal eigenvalue
approximation (\ref{eq:rho_ell}), which predicts a different limit at
short times and small $\ell$.  This difference is seen on panel {\bf
(a)} of Fig. \ref{fig:rho_comparison}.  A better understanding of the
origins of this deviation presents an interesting open problem.

\end{document}